\documentclass[fleqn,usenatbib]{mnras}

\usepackage{newtxtext,newtxmath}

\usepackage[T1]{fontenc}

\DeclareRobustCommand{\VAN}[3]{#2}
\let\VANthebibliography\thebibliography
\def\thebibliography{\DeclareRobustCommand{\VAN}[3]{##3}\VANthebibliography}

\usepackage{graphicx}	
\usepackage{amsmath}	
\usepackage{cleveref}

\newcommand{\maxsmooth}{\texttt{maxsmooth}}
\newcommand{\cvxopt}{\texttt{CVXOPT}}
\newcommand{\multinest}{\texttt{MultiNest}}
\newcommand{\scipy}{\texttt{Scipy}}
\crefformat{figure}{Fig.~#2#1#3}
\crefformat{equation}{equation~(#2#1#3)}
\crefformat{section}{section~#2#1#3}
\crefformat{table}{Tab.~#2#1#3}
\crefformat{appendix}{App.~#2#1#3}
\crefmultiformat{figure}{Figs.~#2#1#3}{~and~#2#1#3}%
    {,~#2#1#3}{,~#2#1#3}
\crefrangeformat{figure}{Figs.~(#3#1#4--#5#2#6)}

\title[\maxsmooth: rapid maximally smooth function fitting]{\maxsmooth: rapid maximally smooth function fitting with applications in Global 21-cm cosmology}

\author[H. T. J. Bevins et al.]{H. T. J. Bevins$^{1}$\thanks{E-mail: htjb2@cam.ac.uk},
    W. J. Handley$^{1, 2}$,
    A. Fialkov$^{2, 3}$,
    E. de Lera Acedo$^{1, 2}$,
    L. J. Greenhill$^{4}$,
    \newauthor
    D. C. Price$^{4, 5}$
    \\
    $^{1}$Astrophysics Group, Cavendish Laboratory, J. J. Thomson Avenue, Cambridge, CB3 0HE, UK\\
    $^{2}$Kavli Institute for Cosmology, Madingley Road, Cambridge CB3 0HA, UK \\
    $^{3}$Institute of Astronomy, University of Cambridge, Madingley Road, Cambridge CB3 0HA, UK\\
    $^{4}$Harvard-Smithsonian Center for Astrophysics, MS42, 60 Garden Street, Cambridge MA 02138 USA \\
    $^{5}$Centre for Astrophysics and Supercomputing, Swinburne University of Technology, Hawthorn VIC 3122 Australia
}

\date{Accepted XXX. Received YYY; in original form ZZZ}

\pubyear{2020}

\begin{document}
\label{firstpage}
\pagerange{\pageref{firstpage}--\pageref{lastpage}}
\maketitle

\begin{abstract}
Maximally Smooth Functions~(MSFs) are a form of constrained functions in which there are no inflection points or zero crossings in high order derivatives. Consequently, they have applications to signal recovery in experiments where signals of interest are expected to be non-smooth features masked by larger smooth signals or foregrounds. They can also act as a powerful tool for diagnosing the presence of systematics. The constrained nature of MSFs makes fitting these functions a non-trivial task. We introduce \maxsmooth, an open source package that uses quadratic programming to rapidly fit MSFs. We demonstrate the efficiency and reliability of \maxsmooth~by comparison to commonly used fitting routines and show that we can reduce the fitting time by approximately two orders of magnitude. We introduce and implement with \maxsmooth~Partially Smooth Functions, which are useful for describing elements of non-smooth structure in foregrounds. This work has been motivated by the problem of foreground modelling in 21-cm cosmology. We discuss applications of \maxsmooth~to 21-cm cosmology and highlight this with examples using data from the Experiment to Detect the Global Epoch of Reionization Signature~(EDGES) and the Large-aperture Experiment to Detect the Dark Ages~(LEDA) experiments. We demonstrate the presence of a sinusoidal systematic in the EDGES data with a log-evidence difference of $86.19\pm0.12$ when compared to a pure foreground fit. MSFs are applied to data from LEDA for the first time in this paper and we identify the presence of sinusoidal systematics. \maxsmooth~is pip installable and available for download at: \url{https://github.com/htjb/maxsmooth}
\end{abstract}




\section{Introduction}
\label{sec:intro}

Maximally Smooth Functions~(MSFs), functions with no inflection points or zero crossings in higher order derivatives, were first proposed by \cite{MSFRE} for modelling foregrounds in experiments to detect spectral signatures from the Epoch of Recombination. They are designed for modelling smooth structures in experimental data that are several orders of magnitude larger than non-smooth signals of interest and to leave behind signals and, where present, systematics in residuals \citep[see also,][]{MSFCD}. MSFs can be considered part of a family of functions, which we refer to as Derivative Constrained Functions~(DCFs) and includes functions with no turning points, Completely Smooth Functions~(CSFs) and functions with a select number of non-zero crossing high order derivatives, Partially Smooth Functions~(PSFs). We refer to the high magnitude smooth components of the data as foregrounds throughout this paper.

Our primary focus here is the application of DCFs to the field of Global 21-cm cosmology. \cite{Shaver1999} suggested that reionization of neutral hydrogen in the early universe would result in a sharp step in the global spectrum of the sky, and that this signal should be separable from the smooth spectrum emission that dominates the sky temperature at radio wavelengths, $70 - 240$~MHz. \cite{Pritchard2010} showed that the foreground emission, if modelled using a low-order polynomial, could be subtracted from the global sky spectrum to retrieve signals from the Epoch of Reionization~(EoR) of order $100$~mK; \cite{Harker2012} and \cite{Bernardi2015} expounded on this work, including further instrumental effects.

In comparison to unconstrained polynomials, DCFs are better able to separate the smooth foreground spectra from the anticipated EoR signals and instrumental systematics \citep{MSFCD}. This motivates their use in Global 21-cm cosmology experiments such as; REACH \citep[Radio Experiment for the Analysis of Cosmic Hydrogen,][]{REACH}, SARAS \citep[Shaped Antenna measurement of the background RAdio Spectrum,][]{SARAS}, EDGES \citep[Experiment to Detect the Global Epoch of Reionization Signature,][]{EDGES_LB}, LEDA \citep[Large-aperture Experiment to Detect the Dark Ages,][]{LEDA}, PRIZM \citep[Probing Radio Intensity at High-Z from Marion,][]{Prizm}, BIGHORNS \citep[Broadband Instrument for Global HydrOgen ReioNisation Signal][]{BIGHORNS}, SCI-HI \citep[Sonda Cosmol\'ogica de las Islas para la Detecci\'on de Hidr\'ogeno Neutro][]{SCIHI} and MIST~(Mapper of the IGM Spin Temperature, \url{http://www.physics.mcgill.ca/mist/}).

Specifically, the Global 21-cm signal is the sky averaged temperature deviation between the cosmic microwave background~(CMB) and the spin temperature of hydrogen gas during the EoR and the period of cosmic history known as the Cosmic Dawn~(CD). The physics of the Global 21-cm signal have been extensively reviewed and what follows is a brief summary of the structure-defining processes. For further details see \cite{Furlanetto2006}, \cite{Pritchard2012} and \cite{Barkana2016}.

During the CD the first stars begin to form in halos that have accumulated mass under gravity. In the following EoR the neutral hydrogen gas becomes completely ionised by the ultraviolet emission from the first luminous sources. The structure of the 21-cm signal is defined by various astrophysical processes including the adiabatic cooling and collisional coupling of the neutral hydrogen and the gas, the Wouthuysen-Field~(WF) effect, X-ray heating and ionization.

At high redshifts, during the dark ages of the universe, collisions between neutral hydrogen, other hydrogen atoms, electrons and protons couple the spin temperature to the gas temperature. The gas cools adiabatically and at a faster rate than the CMB producing an absorption against the CMB before the first stars begin to form. As the universe expands the density of baryons reduces and collisional coupling becomes inefficient. The spin temperature is driven consequently back to the CMB temperature by radiative coupling.

Once the first luminous sources begin to form in the CD, the WF effect begins to become important. The absorption and re-emission of Ly-$\alpha$ photons from the first luminous sources by neutral hydrogen causes spin flip transitions and drives the distribution of hydrogen atoms in the excited and the ground 21-cm states  \citep[]{Wouthuysen, Field}. As a result, the spin temperature couples to the gas temperature again, which has continued to cool adiabatically producing another absorption trough against the CMB.

X-ray sources heat the gas at later times and, if sufficient heating occurs, this causes the gas temperature to exceed the CMB temperature producing an emission above the CMB. The primary sources of X-ray emission during this epoch are thought to be X-ray binaries \citep[e.g.][]{Fragos2013}. Finally, at lower redshifts the neutral hydrogen gas is ionized by UV emission. As the gas becomes completely ionized the signal disappears against the background of the CMB. The CD, the subsequent reheating and reionization of the gas are the focus of the experiments listed above.

The above processes do not occur at independent epochs and do not start and stop instantaneously. Consequently, the structure of the signal is determined by the interplay between these mechanisms and by the change in the dominant processes with time. The exact timing and intensity  of the signal is only broadly understood within a theoretical parameter space \citep[]{Cohen, SARAS_Const, EDGES-HB, Cohen2020}. Experiments that search for the Global 21-cm signal are attempting to detect a signal, according to standard $\Lambda$CDM cosmology, approximately $250$~mK in foregrounds of up to $10^4~-~10^5$ times brighter. 

These high-magnitude foregrounds are dominated by synchrotron and free-free emission in the Galaxy and extragalactic radio sources which have smooth power law structures. Modelling of these foregrounds without signal loss is essential for an accurate detection and not always possible with unconstrained polynomials. However, unconstrained polynomials and linear combinations of unconstrained polynomials remain the traditionally used foreground model in 21-cm experiments \citep{EDGES_LB, SARAS_Const, EDGES-HB}.

The bandwidth is determined by the intrinsic frequency of the 21-cm transition, $1420.4$~MHz, which is redshifted by the expansion of the universe. Studies of Gunn-Peterson troughs in quasar spectra and of the CMB anisotropies put the end of the EoR at $z\approx6$ \citep{Becker2001, Spergel2007, Planck2018}. It is predicted that the onset of star formation occurred at $z\sim 30$ \citep{Abel2002} and consequently the bandwidth of interest for 21-cm cosmology is approximately $50 - 200$~MHz.

\begin{figure}
    \centering
    \includegraphics{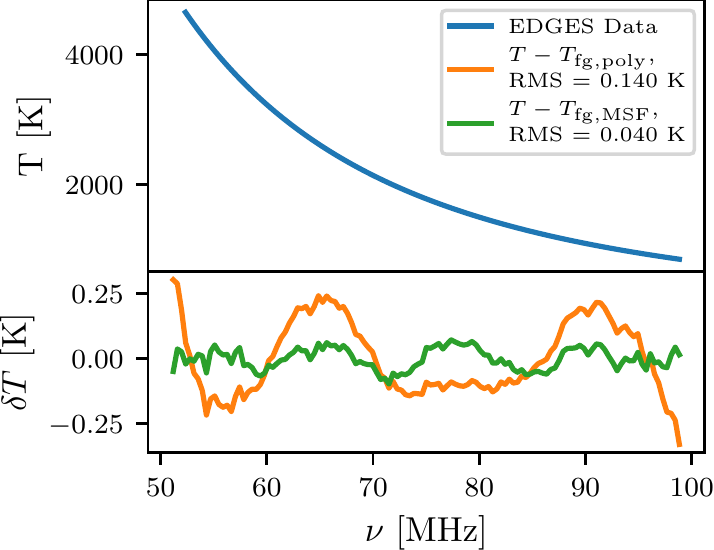}
    \caption{An example of the abilities of MSFs and \maxsmooth~using the publicly available Global 21-cm EDGES low-band experiment data. The top panel shows the EDGES data, blue, and the bottom panel shows the residuals after fitting and removing an unconstrained polynomial, orange, and an MSF, green. The MSF fits the data to a higher degree of accuracy and reveals a systematic that has been partially removed by the polynomial as part of the foreground. The polynomial is given by equation~(2) of \protect\cite{EDGES_LB} and is taken to be 5\textsuperscript{th} order. We use the best fitting 11\textsuperscript{th} order MSF from the built-in library in \maxsmooth~to illustrate the quality of fit recovered.}
    \label{fig:fig0}
\end{figure}

\Cref{fig:fig0} shows an example of the application of MSFs to 21-cm cosmology. Here we have fitted publicly available data from the EDGES low band experiment with an MSF and a 5\textsuperscript{th} order polynomial of the form given by equation~(2) in \cite{EDGES_LB}. The MSF is shown to fit the foreground to a higher degree of precision and potentially reveal a sinusoidal systematic which has been previously identified in the data~\citep{Hills, MSF-EDGES, Sims}. A more detailed discussion of the EDGES data can be found in \cref{sec:EDGES_fits}.

The constrained nature of DCFs, namely that specific derivatives do not cross zero in the domain of interest, makes fitting these functions a non-trivial task. While this has been historically performed with optimization routines such as Basin-hopping \citep[][]{Basinhopping} and Nelder-Mead \citep[][]{Nelder-Mead} we find that the use of quadratic programming \citep[][]{qp} is considerably more computationally efficient and reliable. Our DCF code, \maxsmooth~is therefore based on quadratic programming and uses the Python based convex optimisation code, \cvxopt~\citep[][]{cvxopt}. A discussion of quadratic programming can be found in \cref{app:qp}. 

The constraints on a DCF are not explicitly linear but are piecewise linear with various combinations of positive and negative signs on the high order derivatives. For low order, $N$, DCFs testing every combination of positive and negative signs is a computationally inexpensive task. However, this becomes increasingly time consuming with increasing $N$ and \maxsmooth~uses a cascading routine in combination with a directional exploration to quickly search the discrete sign spaces.

DCFs can be formed from a variety of different basis functions and \maxsmooth~has a built-in library. The library is not intended to be complete, and the user can implement their own basis functions. For basis functions in which the number of high order derivatives is not finite, \maxsmooth~constrains derivatives up to order $m = N - 2$. MSFs form the basis of the analysis performed in this paper and we focus on their uses and applications. However, the description of MSFs can be more broadly applied to DCFs.

In \cref{sec:MSFs} we describe MSFs in more detail and give examples using the built-in DCFs in \maxsmooth. In \cref{sec:qp} we discuss the application of quadratic programming to DCF fitting with reference to \cvxopt~and the piecewise linear constraints on the derivatives. \Cref{sec:Eff} discusses the fitting algorithm implemented by \maxsmooth~and compares its efficiency to alternative optimization routines. In \cref{sec:21} we discuss the use of PSFs in 21-cm cosmology. This discussion is then followed by the application of the fitting routine to the EDGES low band data \citep{EDGES_LB} and data from LEDA \citep{LEDA}. We conclude in \cref{sec:conclusions} highlighting the particular applications of \maxsmooth.

\section{Maximally Smooth Functions}
\label{sec:MSFs}

MSFs are functions which feature no inflection points or zero crossings in higher order derivatives \citep[see][]{MSFRE, MSFCD}. The coefficients of the basis functions are constrained such that the $m$\textsuperscript{th} order derivative satisfies
\begin{equation}
    \frac{d^my}{dx^m}~\geq0~~\textnormal{or}~~  \frac{d^my}{dx^m}~\leq0,
    \label{eq:gen_cond}
\end{equation}
where $x$ and $y$ define the independent and dependent variables and for MSFs $m~\ge~2$. More generally for DCFs $m$ can be greater or equal to any value or equal to a select set of derivative orders. \maxsmooth~features seven built-in DCFs which we use for fitting. Their functional forms and derivatives are shown in \cref{tab:basis_functions}.

\begin{table*}
    \begin{tabular}{ccc}
        \hline
          Name & Function & Derivatives \\
         \hline
         Normalised Polynomial 
         & \parbox{4cm}{\begin{equation}
             y~=~y_0\sum_{k=0}^{N}~a_{k}~\bigg(\frac{x}{x_0}\bigg)^k
             \label{eq:norm_poly}
         \end{equation}}
         & \parbox{8cm}{\begin{equation*}
            \frac{d^my}{dx^m}~=~y_0~\sum_{k=0}^{N-m}~\frac{(m~+~k)!}{k!}~a_{m+k}~\bigg(\frac{x^k}{x_0^{m+k}}\bigg)
        \end{equation*}} \\
         Polynomial 
         & \parbox{4cm}{\begin{equation}
            y~=~\sum_{k=0}^{N}~a_{k}~x^k \label{eq:additional_basis_a}
         \end{equation}}
         & \parbox{8cm}{\begin{equation*}
            \frac{d^my}{dx^m}~=~\sum_{k=0}^{N-m}~\frac{(m~+~k)!}{k!}~a_{m+k}~x^k
         \end{equation*}} \\
         Difference Polynomial 
         & \parbox{4cm}{\begin{equation}
            y~=~\sum_{k=0}^{N}~a_{k} (x~-~x_0)^k
            \label{eq:additional_basis_b}
         \end{equation}}
         & \parbox{8cm}{\begin{equation*}
            \frac{d^my}{dx^m}~=~\sum_{k=0}^{N-m}~\frac{(m~+~k)!}{k!}~a_{m+k}~(x~-~x_0)^k
         \end{equation*}}\\
         Log Polynomial 
         &  \parbox{4cm}{\begin{equation}
            y~=~\sum_{k=0}^{N}~a_{k}~\log_{10}\bigg(\frac{x}{x_0}\bigg)^k 
            \label{eq:log_poly}
         \end{equation}}
         &  \parbox{8cm}{\begin{equation*}
            \frac{d^my}{d\log_{10}(x/x_0)^m}~=~\sum_{k=0}^{N-m}~\frac{(m~+~k)!}{k!}~a_{m+k}~\log_{10}\bigg(\frac{x}{x_0}\bigg)^k
         \end{equation*}} \\
         Log Log Polynomial 
         & \parbox{4cm}{\begin{equation}
            y~=~10^{\sum_{k=0}^{N}~a_{k}~\log_{10}(x)^k}
            \label{eq:loglog_poly}
         \end{equation}}
         & \parbox{8cm}{\begin{equation*}
            \frac{d^m\log_{10}(y)}{d\log_{10}(x)^m}~=~\sum_{k=0}^{N-m}~\frac{(m~+~k)!}{k!}~a_{m+k}~\log_{10}(x)^k
         \end{equation*}} \\
         Legendre & \parbox{4cm}{\begin{equation}
            y~=~\sum_{k=0}^{N}~a_{k} P_k(z)
            \label{eq:legendre}
         \end{equation}}
         &  \parbox{8cm}{\begin{equation*} 
            \frac{d^my}{dz^m} = \sum_{k=0}^{N-m}~\frac{(-1)^m~P^m_k(z)}{(1-z^2)^{\frac{m}{2}}}
            \end{equation*}} \\
         Exponential 
         & \parbox{4cm}{\begin{equation}
            y~=~y_0\sum_{k=0}^{N}~a_{k}~\exp\bigg(-k~\frac{x}{x_0}\bigg)
         \label{eq:exponential}
         \end{equation}}
         & \parbox{8cm}{\begin{equation*}
            \frac{d^my}{dx^m} = y_0~\sum_{k=0}^{N} \bigg(\frac{-k}{x_0}\bigg)^m a_k \exp\bigg(-k~\frac{x}{x_0}\bigg)
         \end{equation*}}\\
    \end{tabular}
    \caption{The DCF models built-in to \maxsmooth~along with expressions for their $m$\textsuperscript{th} order derivatives. For all functions $y_0$ and $x_0$ are pivot points in the data sets. More details on each DCF function can be found in the text.}
    \label{tab:basis_functions}
\end{table*}

Generally, the DCF functions can be decomposed in terms of basis functions, $\phi$ and parameters, $a_k$ as
\begin{equation}
    y~=~\sum_{k=0}^{N}~a_{k}~\phi_k(x).
    \label{eq:general_poly}
\end{equation}
For the first DCF shown in \cref{tab:basis_functions}, the Normalised Polynomial model, the basis functions are given by
\begin{equation}
    \phi_k(x)~=~y_0~\bigg(\frac{x}{x_0}\bigg)^k,
\end{equation} 
where $y_0$ and $x_0$ correspond to a pivot point, defaulted to the mid-point, in the data sets. The normalised nature of this polynomial model ensures that the fit parameters, $a_k$, are of order unity. Here $N$ is the order of the DCF and can take on any value. However, for a given model and data set there is a limiting value beyond which a further increase in $N$ does not increase the quality of the fit and this is illustrated in \cref{sec:EDGES_fits} and \cref{fig:EDGES_basis}. The DCF model will have powers from 0 to $N-1$.

Two more basis functions built-in to \maxsmooth~are given by the Polynomial and Difference Polynomial models where the latter is based on the basis function used in \cite{MSFCD}. The built-in set of models is not meant to be complete with the intention for it to be extended in the future.

The fourth basis function built-in to \maxsmooth, Log Polynomial, produces an MSF in $y - \log_{10}(x)$ space. \maxsmooth~is also capable of fitting a DCF in $\log_{10}(y) - \log_{10}(x)$ space given in \cref{tab:basis_functions} as the Log Log Polynomial model. In this instance the function is constrained by derivatives in $\log_{10}(y) - \log_{10}(x)$ space. This can be advantageous in situations where the foregrounds are expected to take on a power law structure.

The penultimate basis function in the \maxsmooth~library of models is built from the orthogonal Legendre Polynomials, $P_k(z)$, where z is a variable of length $y$ over the range [-1, 1]. The $m$\textsuperscript{th} order derivatives of this model are determined by the Associated Legendre Polynomials, $P^m_k(z)$. By definition the Legendre polynomials are a linear combination of the basis functions of the Normalised Polynomial model. This is true also for the Polynomial model and less trivially for the Difference Polynomial model.

Typically the basis functions are designed so that \cref{eq:general_poly} has a finite number of high order derivatives and is consequently polynomial in nature. However, more elaborate models with an infinite number of derivatives are plausible if we consider these functions to be maximally smooth when all derivatives with $ 2\leq m \leq N-2$ are constrained. The final DCF model built-in to \maxsmooth, the Exponential DCF, is an example of this with exponential basis functions. This model fails at high $N$ where the exponential cannot be computationally calculated. However, it is a useful example of a DCF with infinite derivatives and performs well with low values of $N$. The exact value of $N$ at which this basis begins to fail is determined by the magnitude of the $x$ data. An alternative example of a basis with infinite derivatives that \maxsmooth~is capable of fitting would be a polynomial function with non-integer powers.

Generally, the form of the basis function is important in determining the quality of the residuals and careful exploration of the basis functions are needed in order to draw sensible conclusions about the data set. Again, this is illustrated with an example in \cref{sec:EDGES_fits}. We also note that DCFs fitted in $y - \log_{10}(x)$ space, $\log_{10}(y) - \log_{10}(x)$ space, $y - x$ space or $y - z$ space are not equivalent since the form of the constraints and the function that we minimise are different in each case. This is discussed further in \cref{sec:qp}.

With appropriate normalisation \maxsmooth~will be able to transform any basis function into a `standard' form, which can be solved easily and transformed back into the initial basis function choice. Designing and automating such a normalisation is the subject of ongoing work. Provided this `standard' form is chosen well such that it will always return the best quality fits and is computationally solvable with quadratic programming, the initial choice of basis function will largely be negated. Its form will only be determined by the need of the user to model their foreground using a specific model. For example, in 21-cm cosmology this specific model may be a linearised physical model of the data fitted as an MSF. While normalisation remains absent in \maxsmooth, the user has the ability to input normalised $x$ and $y$ data.

For quadratic programming, the method used here to fit MSFs, it is useful to reformulate \cref{eq:general_poly} as a matrix equation. Explicitly we have discrete data points $y_i$ and $x_i$ which means that $\phi_k(\mathbf{x})$ forms a two dimensional matrix, $\mathbf{\Phi}$. The matrix of basis functions has dimensions $(D\times N)$ where D is the length of $\mathbf{y}$ and $N$, as before, is the order of the function. We write this as
\begin{equation}
    \begin{bmatrix}
    y_0 \\ \vdots \\ y_D
    \end{bmatrix}
    =
    \begin{bmatrix}
    \phi_{00} & \dots & \phi_{0(N-1)} \\
    \vdots & \ddots & \\
    \phi_{D0} & & \phi_{D(N-1)}
    \end{bmatrix}
    \begin{bmatrix}
    a_0 \\ \vdots \\ a_{(N-1)}
    \end{bmatrix},
\end{equation}
where $\phi_{ik} = \phi_k(x_i)$. We can summarise this as
\begin{equation}
    \mathbf{y}~=~\mathbf{\Phi}~\mathbf{a},
    \label{eq:gen_poly_matrix}
\end{equation}
where $\mathbf{a}$ is a column vector of length $N$ representing the parameters. For the polynomial basis function in \cref{eq:additional_basis_a}, the element $\phi_{D0}$, or $\phi_0(x_D)$, has the form $x_D^0$ and $\phi_{0(N-1)}$ has the form $x_0^{(N-1)}$.

Reformulating \cref{eq:gen_cond} in terms of matrices for quadratic programming is more complicated. If we take the definition of the condition with the derivative $\leq 0$ and write this in the form of a matrix for a given derivative order $m$ we find
\begin{equation}
    \begin{bmatrix}
    \frac{d^my}{dx^m}_0 \\
    \vdots \\
    \frac{d^my}{dx^m}_{D} \\
    \end{bmatrix}
    \leq
    \begin{bmatrix}
    0 \\ \vdots \\ 0
    \end{bmatrix},
    \label{eq:derivative_matrices}
\end{equation}
where both matrices are columns of length $D$. Each row in the derivative matrix corresponds to an evaluation of the $m$\textsuperscript{th} order derivative for a given $y_i$ and $x_i$.

We can expand the elements of the derivative matrix out into a matrix of derivative prefactors, $\mathbf{G}$, and the matrix of parameters $\mathbf{a}$, as in \cref{eq:gen_poly_matrix}, which is useful for implementing quadratic programming. This is best illustrated with an example, we will look at the simple case of $N~=~3$ with one constrained derivative $m~=~2$. We will say that our data sets have a length $D~=~4$ and choose the simplest functional form for our MSF given by \cref{eq:additional_basis_a}. In this case $G$ is given by
\begin{equation}
    G_k^m(x_i) = \frac{(m~+~k)!}{k!}~x_i^k,
    \label{eq:prefactors}
\end{equation}
for the range $k = 0$ up to but not including $N-m$. For this problem $k$ has only one value $0$ which would produce a column matrix of elements of length $D$. However, since we need to multiply this by the column matrix $\mathbf{a}$ of length $N$ then the matrix $\mathbf{G}$ should have dimensions of $D \times N$. The additional elements in this instance are $0$ so that the product of these elements with the corresponding elements of $\mathbf{a}$ equals $0$. For example here the evaluation of the second order derivative $\mathbf{Ga}$ for the first data element will be
\begin{equation}
    \frac{d^2y}{dx^2}_0 = a_0 0 + a_1 0 + a_2 \frac{(2~+~0)!}{0!}~(x_0)^0.
\end{equation}

Generally if the row elements of $\mathbf{G}$ have a position from $0 - (N-1)$ then the elements with position $\leq m - 1$ will be $0$. The matrix $\mathbf{G}$ for our specific problem then becomes
\begin{equation}
    \mathbf{G} =
    \begin{bmatrix}
    0 & 0 & G_0^2(x_0) \\
    0 & 0 & G_0^2(x_1) \\
    0 & 0 & G_0^2(x_2) \\
    0 & 0 & G_0^2(x_3) \\
    \end{bmatrix},
\end{equation}
which when multiplied by $\mathbf{a}$ gives us the evaluation of the derivatives as a column matrix with length $D = 4$.

For quadratic programming we need one matrix expression for all of the constraints on our function. Our definition of $\mathbf{G}$ scales with the order of the DCF so that for $N~=~4$ we have
\begin{equation}
    \mathbf{G} =
    \begin{bmatrix}
    0 & 0 & G_0^2(x_0) & G_1^2(x_0)\\
    0 & 0 & G_0^2(x_1) & G_1^2(x_1)\\
    0 & 0 & G_0^2(x_2) & G_1^2(x_2)\\
    0 & 0 & G_0^2(x_3) & G_1^2(x_3)\\
    0 & 0 & 0 & G_0^3(x_0)\\
    0 & 0 & 0 & G_0^3(x_1)\\
    0 & 0 & 0 & G_0^3(x_2)\\
    0 & 0 & 0 & G_0^3(x_3)\\
    \end{bmatrix},
    \label{eq:example_G}
\end{equation}
which includes the prefactors on both the $m = 2$ and $m = 3$ derivatives. We can, therefore, re-write \cref{eq:gen_cond} for $\leq 0$ and \cref{eq:derivative_matrices} as
\begin{equation}
    \mathbf{Ga} \leq \mathbf{0},
    \label{eq:cvx_const}
\end{equation}
where generally $\mathbf{G}$ will have shape $(CD) \times N$ and $\mathbf{0}$ will have a length $CD$. Here $C$ is the total number of constrained derivatives and in the two examples above $C = 1$ and $C = 2$ respectively. We can write the first case in \cref{eq:gen_cond} in one of two ways as $\mathbf{Ga} \geq \mathbf{0}$ or as $-(\mathbf{Ga}) \leq \mathbf{0}$. For the implementation of quadratic programming used in \maxsmooth~the second is the most useful and a full discussion of this can be found in \cref{sec:qp}.

In some cases we find that a DCF with one or more high order derivatives free to cross zero is needed to better fit the data. It is to this effect that the potential to allow zero crossings to the fit is built-in to \maxsmooth. However, \maxsmooth~will not force zero crossings and produce a PSF if it can find a better solution without the need.

\section{Fitting Derivative Constrained Functions Using Quadratic Programming}
\label{sec:qp}

We provide a brief overview of quadratic programming in \cref{app:qp} and what follows is a discussion of the specific problem of fitting DCFs with quadratic programming.

When fitting a curve using the least-squares method we minimize
\begin{equation}
    \chi^2~=~\sum_k^D~(y_{k}~-~y_{\mathrm{fit,}~k})^2,
\end{equation}
where $y_{\mathrm{fit},~k}$ denotes the elements of the fitted model. We can substitute \cref{eq:gen_poly_matrix} for the fitted model and re-write this in terms of matrices as
\begin{equation}
    \chi^2(\mathbf{a})~=~(\mathbf{y}~-~\mathbf{\Phi}~\mathbf{a})^T~(\mathbf{y}~-~\mathbf{\Phi}~\mathbf{a}),
\end{equation}
where we are looking for solutions of the parameters $\mathbf{a}$ that minimise $\chi^2(\mathbf{a})$. When expanded out this becomes
\begin{equation}
    \chi^2(\mathbf{a}) = \mathbf{y}^T\mathbf{y} - 2 \mathbf{y}^T \mathbf{\Phi} \mathbf{a} + \mathbf{a}^T \mathbf{\Phi}^T~\mathbf{\Phi} \mathbf{a}.
\end{equation}
Since $\mathbf{y}^T\mathbf{y}$ is a constant it is irrelevant for the minimisation problem and we can ignore it. We can also divide through by the factor of $2$ and this leaves
\begin{equation}
    \chi^2(\mathbf{a})~=~\frac{1}{2}~\mathbf{a}^T~\mathbf{Q}~\mathbf{a}~+~\mathbf{q}^T~\mathbf{a},
    \label{eq:chi}
\end{equation}
where
\begin{equation}
    \mathbf{Q}~=~ \mathbf{\Phi}^T~\mathbf{\Phi}~~\textnormal{and}~~ \mathbf{q}^T~=~-\mathbf{y}^T~\mathbf{\Phi}.
\end{equation}

As previously discussed in \cref{sec:MSFs}, the constraint in \cref{eq:gen_cond} is not explicitly linear but is two separately testable linear constraints. The quadratic program solver \cvxopt~minimizes \cref{eq:chi} subject to \cref{eq:cvx_const}. It requires $\mathbf{G}$ and $\mathbf{0}$ as inputs which explains the motivation behind defining the stacked matrix of derivatives as $\mathbf{Ga}$. In \cvxopt~the identity is fixed and we cannot directly constrain the problem via a greater than or equals inequality.

We can initially force all the derivatives to be positive, the first of the two conditions in \cref{eq:gen_cond}, by multiplying each element in $\mathbf{G}$ by a negative sign as discussed in the previous section. However, it will not necessarily be the case that the optimal DCF fit will have an entirely positive or entirely negative set of derivatives. Rather than forcing the entire matrices to produce positive derivatives we can multiply the elements of $\mathbf{Ga}$ corresponding to given derivatives by a negative sign. Consequently, we have to analyse different discrete sets of sign combinations in order to find the best fit. 

We refer to the combination of signs on $\mathbf{G}$ as the \maxsmooth~signs, $\mathbf{s}$, and we can incorporate this into our definition of $\mathbf{G}$ so that it becomes $\mathbf{G}(\mathbf{s})$. $\mathbf{s}$ is a vector of length $C$ and each element is either given by $1$ for a positive sign or $-1$ for a negative sign. For example in \cref{eq:example_G} since both derivatives are negative the \maxsmooth~signs are $s = [1,~1]$. For an $N$\textsuperscript{th} order MSF there are $N-2$ derivatives with $m~\ge~2$, consequently, there are $2^{(N~-~2)}$ sign combinations. For low order $N$ we can explore this space exhaustively at reasonable computational cost with \cvxopt. However, as $N$ becomes larger, the total number of sign combinations rapidly increases. While $N~=~4$ has $4$ potential sign combinations, we find that $N~=~13$ has $2048$. This would mean performing an exponentially increasing number of \cvxopt~fits which will become increasingly time consuming. An alternative approach navigating through the discrete sign spaces is detailed in \cref{sec:Eff}.

\begin{figure*}
    \centering
    \includegraphics{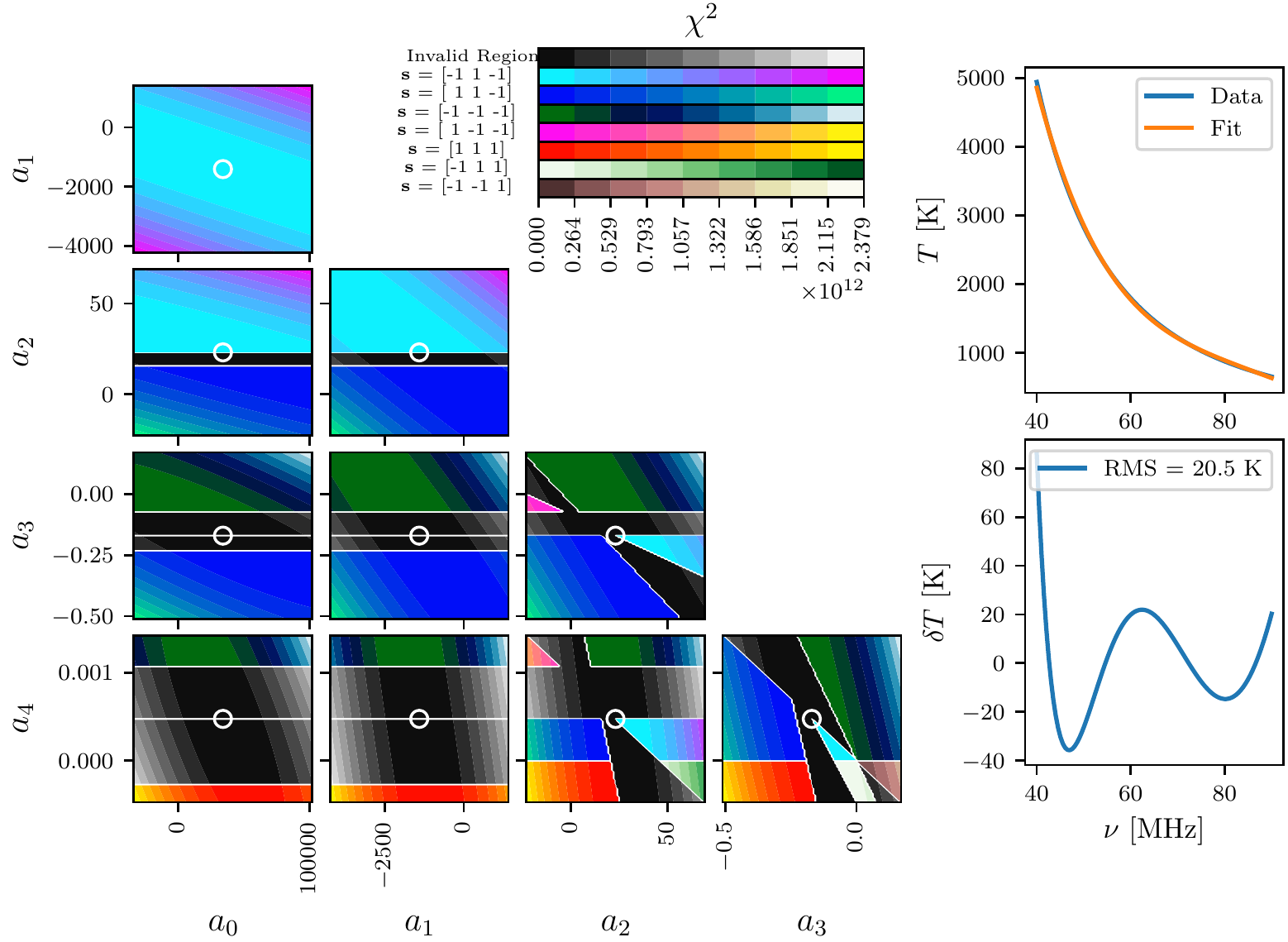}
    \caption{\textbf{Left:} The parameter space of a 5\textsuperscript{th} order MSF of the form given by \cref{eq:additional_basis_a}, fit to data generated with $y = x^{-2.5}$. The tested parameter ranges are taken to be $200\%$ on either side of the optimal results found by \maxsmooth. We maintain the optimal parameter values for three of the parameters when varying the two corresponding to each panel. Black regions of the graph show parameter combinations that violate \cref{eq:gen_cond} and coloured regions correspond to viable regions. The central sampled point, highlighted by the white circles~(at their centers), in each panel corresponds to the optimum parameters and the optimum \maxsmooth~sign combination on the derivatives. Transitions through regions of violation correspond to changes in the sign of one or more high order derivatives. A change in sign of parameter $a_4$ corresponds to a change in the sign of the final constrained derivative. Since this derivative is a constant there is no violated region between these two possible sign combinations. \textbf{Top Right:} The data and the fitted MSF, where $T$ represents the measured sky temperature and $\nu$ is the frequency. \textbf{Bottom Right:} The residuals after subtracting the fitted MSF from the data.}
    \label{fig:poly_params}
\end{figure*}

We can visualise \cref{eq:gen_cond} by varying the parameters of an optimal MSF fit over a given range to get a better understanding of the constraints. In order to perform this analysis we use a simulated noiseless Global 21-cm foreground following $x^{-2.5}$. We perform a 5\textsuperscript{th} order MSF fit with \maxsmooth~on this data using \cref{eq:additional_basis_a} to find the optimal foreground parameters. While this fit will not return the best $\chi^2$, as shown in \cref{fig:poly_params}, it is sufficient to allow us to investigate how variation in the parameters affects the constraints. We vary each parameter's value $200\%$ either side of the optimum found and sample these ranges using 100 points.

\cref{fig:poly_params}, left panel, shows the parameter space for the fit described above. Black regions in the figure are combinations of parameters for which the condition in \cref{eq:gen_cond} is violated. The coloured regions are regions in which the condition is upheld where their colour is related to the \maxsmooth~sign combinations. Each panel in the figure shows the parameter space for two of the five parameters and the contour lines show the values of $\chi^2$ across the parameter ranges. While varying the parameters relevant to each panel we maintain all others at their optimal values found with \maxsmooth. The contour lines help us to determine correlations between the parameters and this is particularly useful when fitting a physically motivated DCF.

Transitions through a region of violation between viable regions correspond to changes in the \maxsmooth~sign of one or more of the derivatives. This is illustrated by the use of different colour maps across the different viable regions. For example, in the panel corresponding to variation in $a_0$ and $a_2$ for $a_2 \leq 15$ $s = 1$ and for $a_2 \geq 25$ $s = -1$ for the $m = 2$ derivative. The transitions become more complex when varying parameters $a_2$, $a_3$ and $a_4$ because these parameters affect the magnitude and signs of multiple high order derivatives. We also see transitions between regions of different sign combinations when $a_4$ switches sign. This causes the final constrained derivative to switch sign because it is a constant multiplied by $a_4$. There is no region of violation between these viable regions because a constant value of $a_4 = 2.5\times 10^{-4}$ meets the MSF constraint as will a constant value of $0$ or $a_4 = -2.5\times 10^{-4}$.

The equivalent graph in 5 dimensions, varying all parameters around their optimal values, would feature 5 dimensional convex faceted regions in which \cref{eq:gen_cond} is met with a unique set of \maxsmooth~signs. This concept scales up and down to higher and lower dimensions of parameter space.

The parameters $a_0$ and $a_1$ do not affect the constrained derivatives of the MSF or the validity of the conditions and the associated colour map gives the optimum \maxsmooth~signs. Since the central sample point of each panel corresponds to the optimum parameters for the fit, this will always be a viable region and will have the same \maxsmooth~sign combination as the panel corresponding to $a_0$ and $a_1$. \cref{fig:poly_params} illustrates this point and, where visible, the central viable sample point always corresponds to derivatives of order $m = 2,~3$ and $4$ having $\mathbf{s} = [-1,~1,~-1]$. 

Where the central sample point in each panel is not visible this is a relic of the sample rate across the parameter space. For example in the panel corresponding to variation of $a_0$ and $a_4$, between the two large regions of violation there is a single value of $a_4$, the optimum, that meets the condition given by \cref{eq:gen_cond}. The tight constraints around the optimum values in the 4 panels in the bottom left corner of the figure are due to the independence of the constraints for an MSF on $a_0$ and $a_1$ and the strong dependence on $a_3$ and $a_4$.

We performed the equivalent fit with the logarithmic basis that has derivatives constrained in $\log_{10}(y) - \log_{10}(x)$ space. The associated parameter graph can be found in \cref{app:param_space_alt_basis}, and a comparison with the graph presented in this section shows that the constraints are much less severe in logarithmic space. The weak constraints mean that all of the discrete sign combinations on the derivatives have similar minimum $\chi^2$ values. This becomes a problem when attempting to quickly search the discrete \maxsmooth~sign spaces and is discussed further in \cref{sec:badproblems}.

Generally, the above conclusion will be specific to the data being fitted here and this analysis is not a complete exploration of the basis functions available. However, since the basis functions in $y - x$ space are all related by linear combinations of each other we find similar parameter distributions for all. Importantly, the analysis highlights the effect that the choice of basis function has on the quality of fit and ease of fitting as well as demonstrating the constrained nature of MSFs. These plots can be produced using \maxsmooth~for any DCF fitting problem.

\section{Navigating Discrete Sign Spaces}
\label{sec:Eff}

This section discusses in more detail the fitting problem, defines the \maxsmooth~algorithm and compares its efficiency with an alternative fitting algorithm. To restate concisely the problem being fitted we have
\begin{equation}
    \begin{split}
        &\min_{a,~s}~~\frac{1}{2}~\mathbf{a}^T~\mathbf{Q}~\mathbf{a}~+~\mathbf{q}^T~\mathbf{a}, \\
        &\mathrm{s.t.}~~\mathbf{G(s)~a} \leq \mathbf{0}.
    \end{split}
\end{equation} 
A `problem' in this context is the combination of the data, order, basis function and constraints on the DCF. 

With \maxsmooth~we can test all possible sign combinations. This is a reliable method and, provided the problem can be solved with quadratic programming, will always give the correct global minimum. When the problem we are interested in is `well defined', we can develop a quicker algorithm that searches or navigates through the discrete \maxsmooth~sign spaces to find the global minimum. Each sign space is a discrete parameter space with its own global minimum as discussed in \cref{sec:qp}. Using quadratic programming on a fit with a specific sign combination will find this global minimum, and we are interested in finding the minimum of these global minima.

A `well defined' problem is one in which the discrete sign spaces have large variance in their minimum $\chi^2$ values and the sign space for the global minimum is easily identifiable. In contrast we can have an `ill defined' problem in which the variance in minimum $\chi^2$ across all sign combinations is small. This concept of `well defined' and `ill defined' problems is explored further in the following two subsections.

\subsection{Well Defined Problems and Discrete Sign Space Searches}
\label{sec:well_defined}

\subsubsection{The $\chi^2$ Distribution}

\begin{figure}
    \centering
    \includegraphics{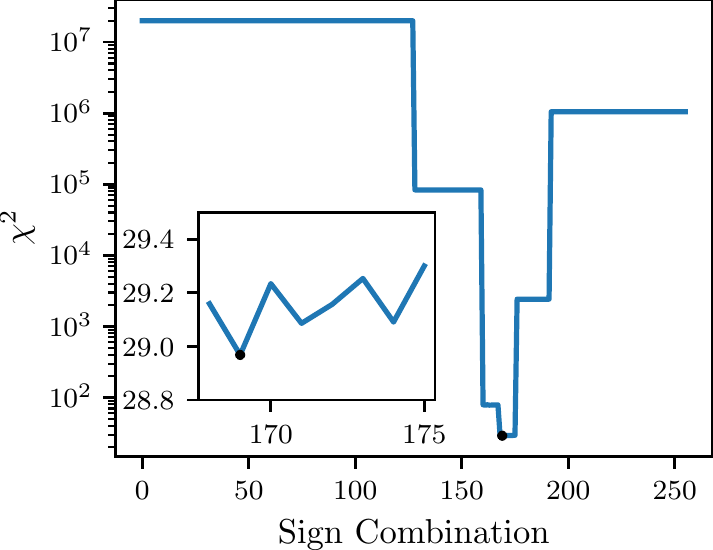}
    \caption{The $\chi^2$ values for the discrete sign combinations on the derivatives for a 10\textsuperscript{th} order MSF fit in $y - \log(x)$ space to a simulated 21-cm foreground. Sign combination 0 corresponds to negative derivatives~(positive \maxsmooth~signs) and 255 corresponds to positive derivatives~(negative \maxsmooth~signs). The signs and sign combination numbers are related by the $(N-2)$ bit binary representation of the number. The global minimum is shown as a single black data point. In the insert, the distribution around the global minimum is shown and here the axis have the same meaning as in the main plot.}
    \label{fig:ChiDistSim21}
\end{figure}

We investigate the distribution of $\chi^2$ values, shown in \cref{fig:ChiDistSim21}, for a 10\textsuperscript{th} order MSF fit of the form given by \cref{eq:log_poly} to a simulated 21-cm foreground, like that shown in \cref{fig:poly_params}. We add Gaussian noise with a standard deviation of $0.5$ to the foreground. For a typical 21-cm experiment this noise is unrealistic and would mask any potential signal in the data, however, it illustrates the behaviour of the \maxsmooth~algorithm when fitting a difficult problem.

In \cref{fig:ChiDistSim21}, a combination of all positive derivatives~(negative signs) and all negative derivatives~(positive signs) corresponds to sign combination numbers 255 and 0 respectively. Specifically, the \maxsmooth~signs, $\mathbf{s}$, are related to the sign combination number by its $C$ bit binary representation, here $C = (N -2)$. In binary the sign combination numbers run from $00000000$ to $11111111$. Each bit represents the sign on the $m$\textsuperscript{th} order derivative with a $1$ representing a negative \maxsmooth~sign. For example, the sign combinations surrounding number 25 are shown in \cref{tab:binary}.

\begin{table}
    \centering
    \begin{tabular}{ccc}
        \hline
         Sign & Binary & \maxsmooth  \\
         Combination &  & Signs, $\mathbf{s}$  \\
         \hline
         23 & 00010111 & [+1, +1, +1, -1, +1, -1, -1, -1]\\
         24 & 00011000 & [+1, +1, +1, -1, -1, +1, +1, +1]\\
         25 & 00011001 & [+1, +1, +1, -1, -1, +1, +1, -1]\\
         26 & 00011010 & [+1, +1, +1, -1, -1, +1, -1, +1]\\
         27 & 00011011 & [+1, +1, +1, -1, -1, +1, -1, -1]\\
    \end{tabular}
    \caption{The table illustrates the relationship between the binary representation of the sign combination number and the \maxsmooth~signs, $\mathbf{s}$. A $1$ in the $(N - 2)$ bit binary representation for an MSF corresponds to a negative \maxsmooth~sign~(positive derivative). The signs and their respective combination numbers are used in the fitting routine and for the visualisation of the $\chi^2$ distribution as shown in \cref{fig:ChiDistSim21}.}
    \label{tab:binary}
\end{table}

Although we note that \cref{fig:poly_params} corresponds to a different problem, we would expect a similar parameter space for the fit performed here. Each region in the equivalent figure would correspond to a single sign combination, and the associated minimum $\chi^2$ value in the regions would give us the data that informs \cref{fig:ChiDistSim21}.

The distribution appears to be composed of smooth steps or shelves; however, when each shelf is studied closer, we find a series of peaks and troughs. This can be seen in the subplot of \cref{fig:ChiDistSim21} which shows the distribution in the neighbourhood of the global minimum found in the large or `global' well. This type of distribution with a large variance in $\chi^2$ is characteristic of a `well defined' problem. We use this example $\chi^2$ distribution to motivate the \maxsmooth~algorithm outlined in the following subsection.

\subsubsection{The \maxsmooth~Sign Navigating Algorithm}

Exploration of the discrete sign spaces for high $N$ can be achieved by exploring the spaces around an iteratively updated optimum sign combination. The \maxsmooth~algorithm begins with a randomly generated set of signs for which the objective function is evaluated and the optimum parameters are found. We flip each individual sign one at a time beginning with the lowest order constrained derivative first. When the objective function is evaluated to be lower than that for the optimum sign combination, we replace it with the new set and repeat the process in a `cascading' routine until the objective function stops decreasing in value.

The local minima shown in \cref{fig:ChiDistSim21} mean that the cascading algorithm is not sufficient to consistently find the global minimum. We can demonstrate this by performing 100 separate runs of the cascading algorithm on the simulated 21-cm foreground, and we use \cref{eq:log_poly} with $N = 10$ to model the MSF as before. We find the true global minimum 79 times and a second local minimum 21 times. For an MSF fit to this simulated date the difference in these local minima is insignificant, $\Delta \chi^2 = 0.12$. However, we see the same behaviour with real data sets from EDGES and LEDA, and when performing joint fits of foregrounds and signals of interest $\Delta \chi^2$ can greatly increase.

The abundance of local minima is determined by the magnitude and presence of signals, systematics and noise in the data. When jointly fitting a signal/systematic model with a DCF foreground, the signal/systematic parameters are estimated by another fitting routine in which \maxsmooth~is wrapped. The initial parameter guess will not be a perfect representation of any real systematic or signal. This, along with a large noise, can produce a large difference between the true foreground and the `foreground' being fitted causing the presence of local minima to become more severe.

To prevent the routine terminating in a local minimum we perform a complete search of the sign spaces surrounding the minimum found after the cascading routine. We refer to this search as a directional exploration and impose limits on its extent. In each direction we limit the number of sign combinations to explore and we limit the maximum allowed increase in $\chi^2$ value. We prevent repeated calculations of the minimum for given signs and treat the minimum of all tested signs as the global minimum.

\begin{figure}
    \centering
    \includegraphics{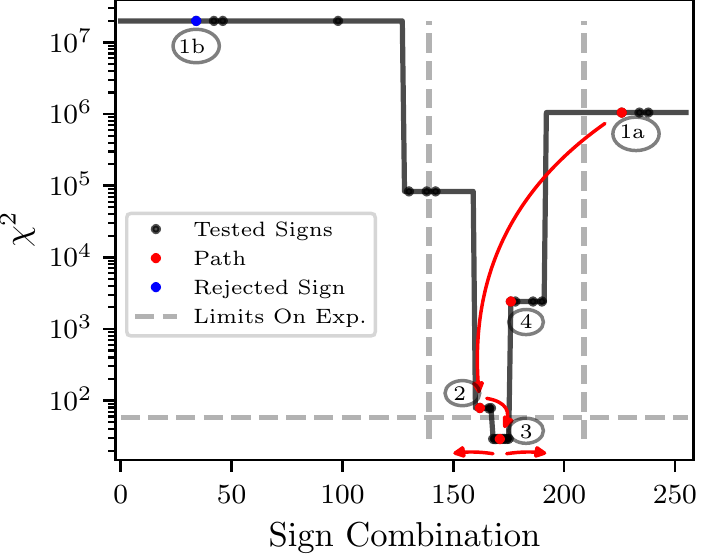}
    \caption{The cascading and directional exploration algorithm in practice against the entire $\chi^2$ distribution for the fit to the simulated 21-cm experiment data. The red arrows show the approximate path of the cascade and directional exploration. The limits on the directional exploration are also shown as dashed grey lines. The point~(1a) shows the initial random starting point and point~(1b) shows a rejected sign combination in the cascade routine from~(1a) to~(2). Point~(2) is an accepted step through the cascade with a $\chi^2$ value smaller than the previous minimum. Point~(3) marks the end of the cascade and the start of the left directional exploration. Finally, point~(4) illustrates the end of the right directional exploration when the $\chi^2$ value exceeds the limit on the directional exploration. The black dots mark the entirety of the searched sign combinations.}
    \label{fig:Run_snapshots}
\end{figure}

We run the consistency test again, with the full \maxsmooth~algorithm, and find that for all 100 trial fits we find the same $\chi^2$ found when testing all sign combinations. In \cref{fig:Run_snapshots} the red arrows show the approximate path taken through the discrete sign spaces against the complete distribution of $\chi^2$. Point~(1a) shows the random starting point in the algorithm, and point ~(1b) shows a rejected sign combination evaluated during the cascade from point~(1a) to~(2). Point~(2), therefore, corresponds to a step through the cascade. Point~(3) marks the end of the cascade and the start of the left directional exploration. Finally, point~(4) shows the end of the right directional exploration where the calculated $\chi^2$ value exceeds the limit on the directional exploration.

The global well tends to be associated with signs that are all positive, all negative or alternating. We see this in \cref{fig:ChiDistSim21} where the minimum falls at sign combination number 169 and number 170, characteristic of the derivatives for the simulated 21-cm foreground, corresponds to alternating positive and negative derivatives from order $m = 2$. Standard patterns of derivative signs can be seen for all data following approximate power laws. All positive derivatives, all negative and alternating signs correspond to data following the approximate power laws $y\approx x^{k}$, $y\approx -x^{k}$, $y\approx x^{-k}$ and $y\approx -x^{-k}$~(see \cref{app:derivatives}). 

The \maxsmooth~algorithm assumes that the global well is present in the $\chi^2$ distribution and this is often the case. The use of DCFs is primarily driven by a desire to constrain previously proposed polynomial models to foregrounds. As a result we would expect that the data being fitted could be described by one of the four approximate power laws highlighted above and that the global minimum will fall around an associated sign combination. In rare cases the global well is not clearly defined and this is described in the following subsection.

\subsection{Ill Defined Problems and their Identification}
\label{sec:badproblems}

\begin{figure}
    \centering
    \includegraphics{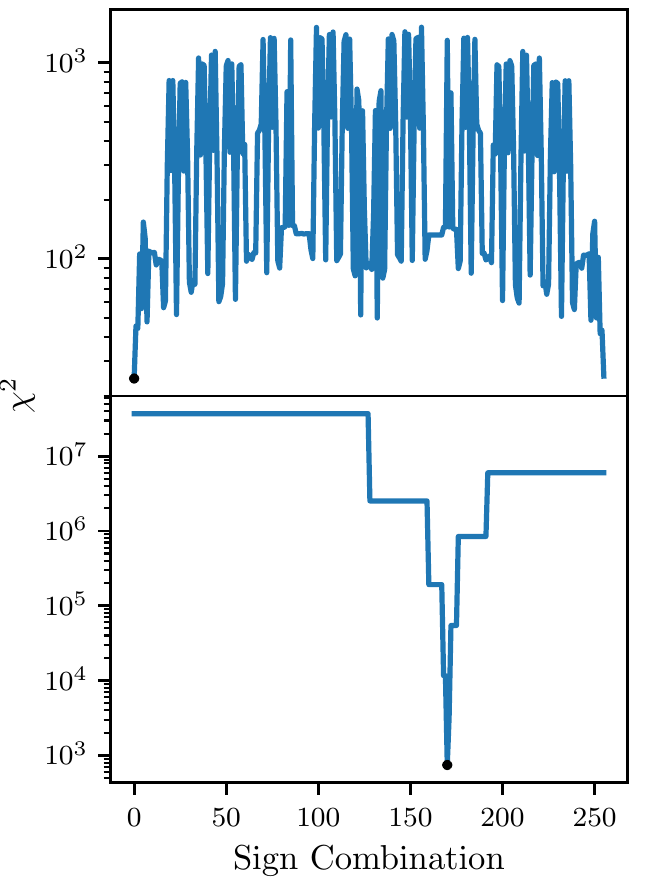}
    \caption{\textbf{Top Panel:} The $\chi^2$ distribution found when fitting simulated 21-cm experiment data with the logarithmic basis function, \cref{eq:loglog_poly}. The distribution has a noise like structure and is difficult to solve with the \maxsmooth~sign navigating algorithm. However, the global minimum can be found by testing all sign combinations with \maxsmooth. The symmetric nature of the distribution stems from the symmetric nature about 0 of the high order derivatives in logarithmic space. \textbf{Bottom Panel:} The same as above using a normalised polynomial given by \cref{eq:norm_poly}. The distribution is clearly defined and easily searchable with the sign navigating routine. The difference between this result and that shown above can be used to understand what makes a problem `ill defined'.}
    \label{fig:bad_problem_chi}
\end{figure}

We can illustrate an `ill defined' problem, with a small variation in $\chi^2$ across the \maxsmooth~sign spaces, by adding a 21-cm signal into the foreground model and fitting this with a 10\textsuperscript{th} order logarithmic MSF defined by \cref{eq:loglog_poly}. We take an example signal model from \cite{Cohen} and add an additional noise of $20$~mK more typical of a 21-cm experiment. The resultant $\chi^2$ distribution with its global minimum is shown in the top panel of \cref{fig:bad_problem_chi}.

The global minimum, shown as a black data point, cannot be found using the \maxsmooth~algorithm. The cascading algorithm may terminate in any of the approximately equal minima and the directional exploration will then quickly terminate because of the limits imposed.

If we repeat the above fit and perform it with \cref{eq:norm_poly} we find that the problem is well defined with a larger $\chi^2$ variation across sign combinations. This is shown in the bottom panel of \cref{fig:bad_problem_chi}. The results, when using \cref{eq:loglog_poly}, are significantly better than when using \cref{eq:norm_poly} meaning it is important to be able to solve `ill defined' problems. This can be done by testing all \maxsmooth~signs but knowing when this is necessary is important if you are expecting to run multiple DCF fits to the same data set. We can focus on diagnosing whether a DCF fit to the data is `ill defined' because a joint fit to the same data set of a DCF and signal of interest will also feature an `ill defined' $\chi^2$ distribution.

We can identify an `ill defined' problem by producing the equivalent of \cref{fig:bad_problem_chi} using \maxsmooth~and visually assessing the $\chi^2$ distribution for a DCF fit. Alternatively, we can use the parameter space plots to identify whether the constraints are weak or not, and if a local minima is returned from the sign navigating routine then the minimum in these plots will appear off centre.

\begin{figure}
    \centering
    \includegraphics{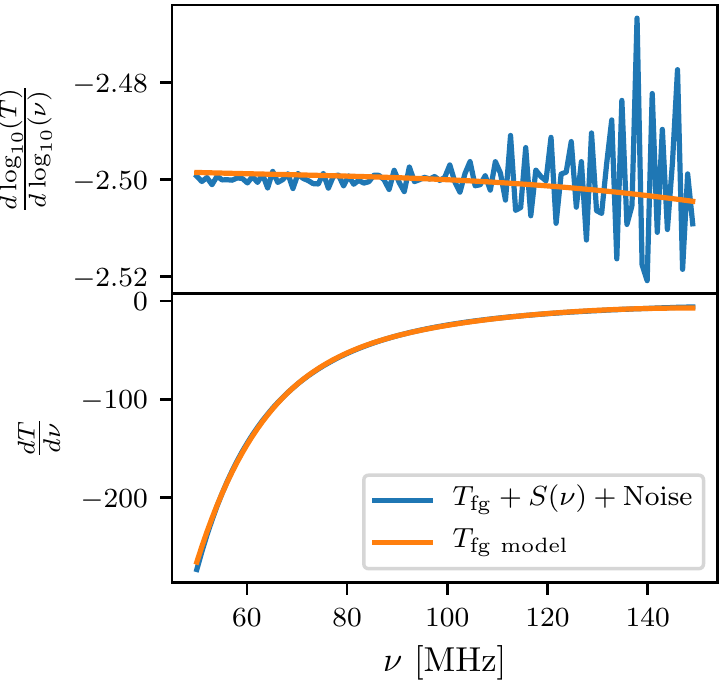}
    \caption{The approximated first derivatives, blue lines, of the mock 21-cm experiment data in logarithmic, top panel, and linear, bottom panel, spaces which correspond to DCF fits performed using \cref{eq:loglog_poly} and \cref{eq:norm_poly} respectively. Also shown as orange lines are the approximate first derivatives of the respective best fits for comparison. The large variance of the first derivative of the data in logarithmic space highlights that there are many different ways to fit this data set `well' producing the $\chi^2$ distribution we see in \cref{fig:bad_problem_chi}.}
    \label{fig:bad_prob_derivatives}
\end{figure}

Assessment of the first derivative of the data can also help to identify an `ill defined' problem. For the example problem this is shown in \cref{fig:bad_prob_derivatives} where the derivatives have been approximated using $\Delta y/ \Delta x$. Higher order derivatives of the data will have similarly complex or simplistic structures in the respective spaces. There are many combinations of parameters that will provide smooth fits with similar $\chi^2$ values in logarithmic space leading to the presence of local minima. This issue will also be present in any data set where the noise or signal of interest are of a similar magnitude to the foreground in $y - x$ space.

\subsection{Comparison With Basin-hopping and Nelder-Mead Methods}
\label{sec:basinhopping}

For comparison of the two methods, testing all sign combinations and navigating through sign spaces, we generate a signal $y$ with polynomial dependence on the coordinate $x$ and a Gaussian random noise with a standard deviation of $0.5$
\begin{equation}
    y~=~0.6~+~2~x~+~4x^3~+~9x^4~+~\textnormal{noise}.
    \label{eq:poly_data}
\end{equation}
We fit this data with a 10\textsuperscript{th} order MSF of the form described by \cref{eq:norm_poly} and assess the $\chi^2$ distribution to find that the problem is well defined. This is as expected since the data follows an approximate $x^k$ power law and we are fitting in linear space.

\begin{figure}
    \centering
    \includegraphics{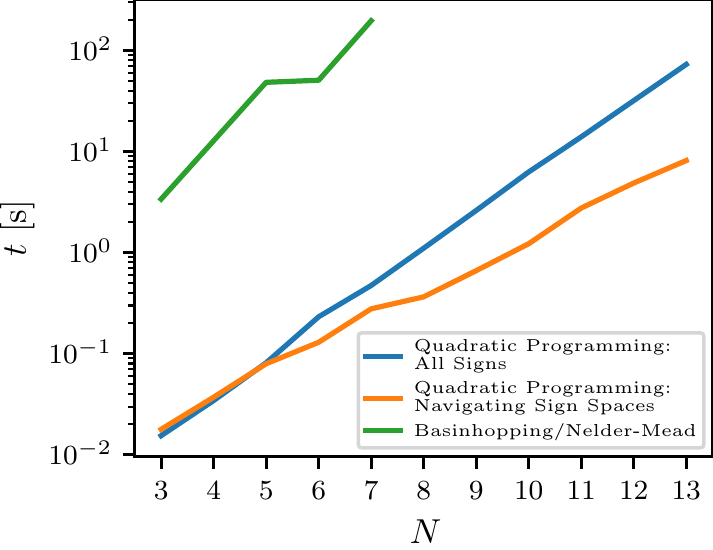}
    \caption{ The time taken by \maxsmooth~to fit MSFs of varying order N to the data described by \cref{eq:poly_data} using the two built-in quadratic programming methods. For comparison the time taken by a Basin-hopping/Nelder-Mead routine is also shown up to $N = 7$ after which the routine fails to find the optimum solutions without adjustments to the routine parameters. All of the fits were performed with \cref{eq:norm_poly} and on the same computing hardware.}
    \label{fig:times}
\end{figure}

The algorithm run time becomes a significant issue when performing joint fits of foregrounds, signals of interest and/or systematics in which multiple DCF fits have to be performed. The time taken to perform both in-built \maxsmooth~routines is shown in \cref{fig:times}. It is quicker to partially sample the available spaces for high $N$ than testing all sign combinations and as discussed for `well defined' problems this will return the minimum $\chi^2$. 

The runtime of the sign navigating routine is dependent on the starting sign combination, the limits imposed on the directional exploration (which is the dominating factor) and the width of the global well. There is no consistent measure of the difference in time taken to fit the data between the two \maxsmooth~methods. However, for the sign navigating routine we are inevitably fitting for a smaller number of the sign combinations than when testing all.

\cref{fig:times} also shows the time taken to fit the data with \cref{eq:norm_poly} using a Basin-hopping routine followed by a Nelder-Mead algorithm, hereafter referred to as BHNM. These two algorithms have been previously used either separately or in conjunction for fitting MSFs \citep{MSFRE, MSFCD, MSF-EDGES}. We find that the BHNM method is approximately 2 orders of magnitude slower than \maxsmooth. Between $N = 3$ and $7$ we find a maximum percentage difference in $\chi^2$ of $\approx 0.04\%$ when comparing the BHNM method with the results from \maxsmooth.

The primary difference in the approaches comes from the division of the parameter space into discrete sign spaces. The BHNM method attempts to search the entire parameter space and penalises parameter combinations that violate \cref{eq:gen_cond}. However, assessment of \cref{fig:poly_params} highlights that this is not a convenient method because across the whole parameter space there are multiple local minima with different sign combinations and transitioning from one 'basin' to another is not trivial for a heavily constrained parameter space. By dividing the space up into discrete sign spaces with \maxsmooth, we can test the entirety of the parameter space, unlike when using the BHNM method, guaranteeing we find the global minimum. We could perform the same division of the space and in each discrete sign space perform a Nelder-Mead or equivalent routine however we use quadratic programming because it is designed specifically for fast and robust constrained optimisation.

For the BHNM method, we show here only fits up to $N=7$ after which it begins to fail without further adjustment of the routine parameters. The freedom to adjust these parameters can be considered a disadvantage that leaves the user to determine whether the routine parameters they have chosen produce a true global minimum. In contrast \maxsmooth~is designed to reliably give the optimum result, with the only adjustable routine parameters being the total number of \cvxopt~iterations and the directional exploration limits.

All of the fits performed in this section were done using the same computing hardware.

\section{Applications for 21-cm Cosmology}
\label{sec:21}

\subsection{The Recovery of Model 21-cm Signals}

A discussion of MSFs and a comparison to unconstrained polynomials for Global 21-cm cosmology can be found in \cite{MSFCD}. Foreground modelling with high order MSFs is shown to accurately recover Global 21-cm signals in simulated data and unconstrained polynomials are shown to introduce additional turning points, when compared to those in a mock signal. The number of additional turning points is shown to increase with polynomial order. The addition of extra turning points can obscure the signal of interest and lead to the false identification of systematics. In the following subsections we look at fitting foregrounds for 21-cm experiments with DCFs and compare these to unconstrained polynomial fits. 

Deviations from a smooth structure can be induced in data by experimental systematics or they can be intrinsic to the foreground. In the case of a smooth foreground, by using PSFs we can correct for non-smooth structure directly with our foreground model rather than separately fitting out systematics. PSFs allow for zero crossings in the high order derivatives but remain more constrained than traditionally used polynomial fits. However, lifting constrains on a DCF model has the potential to also result in signal loss where the level of signal loss is dependent on the presence of non-smooth structure in the foreground and the number of lifted constraints.

In the following analysis the quality of the fit in terms of RMS is of secondary importance. An unconstrained high order polynomial will generally produce residuals with a lower RMS than a DCF. However, a correctly constrained DCF will leave behind the structure of a signal in the residuals meaning it can be identified easily. As a measure of signal structure in the residuals we take the number of turning points, $p$ as used by \cite{MSFCD}. The Global 21-cm signal is expected to have a distinct number of turning points, $2 - 3$, across the bandwidth $~40 - 200$~MHz determined by various astrophysical processes~(see \cref{sec:intro}). The successful application of DCFs in identifying Global signals, being reliant on the presence of non-smooth structure in the data, is therefore bandwidth dependent. However, a comparison of the number of turning points in a mock signal to the residuals after removing a DCF fit from simulated data including the same mock signal will help to identify the degree to which DCFs preserve signals.

\subsubsection{DCFs and 21-cm cosmology}

\begin{figure*}
    \centering
    \includegraphics{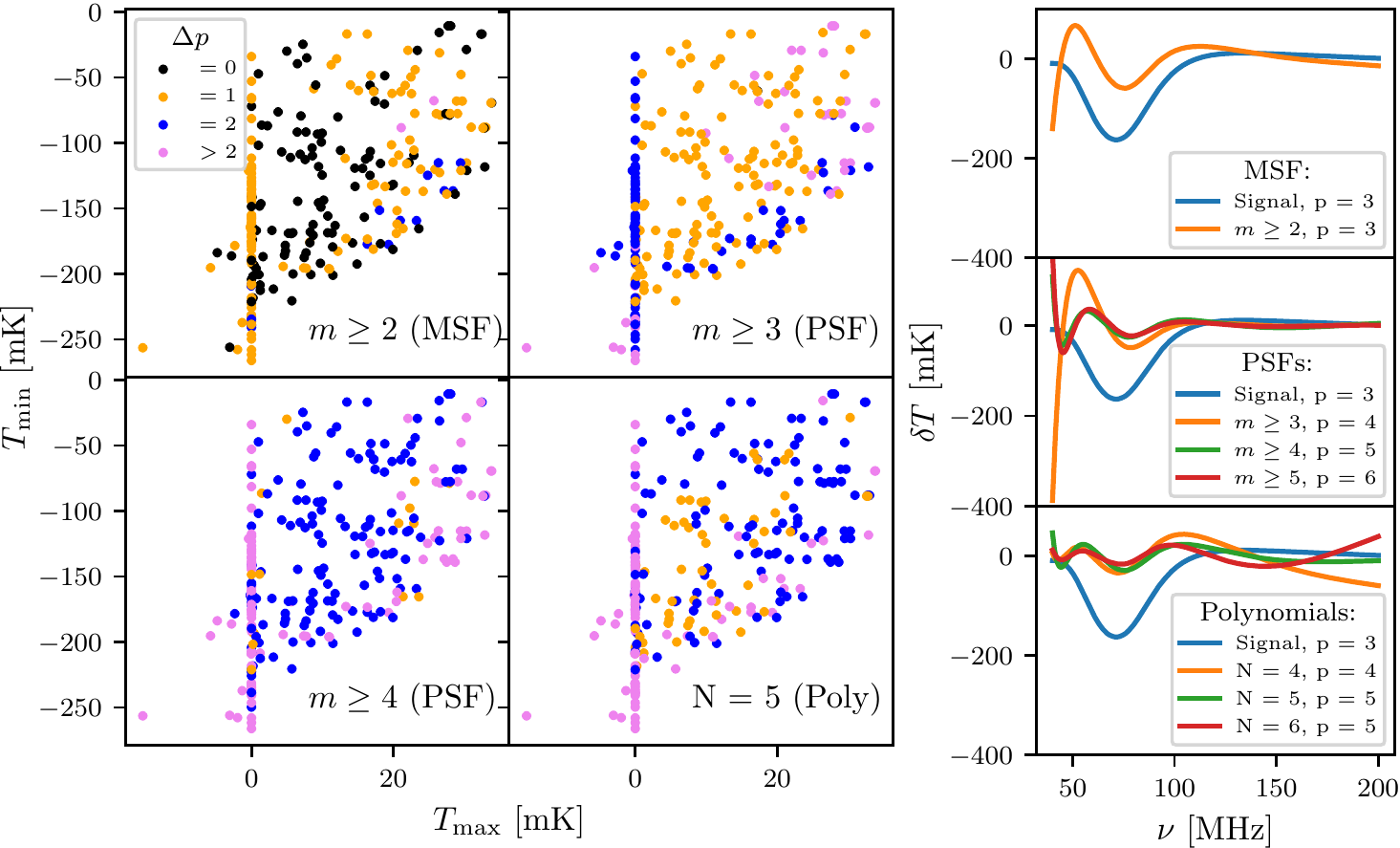}
    \caption{\textbf{Left:} The difference in the number of turning points between the fit residuals and the simulated signal, $\Delta p = p_\mathrm{Fit~Residuals} - p_\mathrm{Signal}$, as a function of the maximum and minimum temperatures of the signal for a smooth foreground model. This is shown for four different foreground models and in each panel the data points correspond to one of the $264$ mock data sets. The graph shows that the MSF fit is the most likely model to return the structure of the signal and that a correctly constrained PSF can more frequently preserve the signal than an unconstrained low order polynomial. It also shows that the fit quality is dependent on the maximum and minimum temperatures of the signal. All of the DCF fits used to produce this graph were logarithmic and 13\textsuperscript{th} order. \textbf{Right:} Shown are examples of how the addition of allowed zero crossings in the high order derivatives of a DCF can affect the residuals and how accurately they preserve the turning points of any signal present. We use a signal model, blue line in all panels, from the theoretically motivated set presented in \protect\cite{Cohen} along with a model of a 21-cm experiment foreground. Fits with an MSF and polynomials are also shown for comparison with the number of turning points, $p$, displayed for each of the residuals~(see legend for details). While the residuals after fitting and removing an MSF do not identically match the signal it is the best representation of the tested foreground models. In this case the residuals represent a smooth baseline subtracted version of the signal as discussed in \citep{MSFCD}.}
    \label{fig:PSF_comp}
\end{figure*}

To compare the performance of DCFs and unconstrained polynomials, we use the sample of 264 signal models, $S(\nu)$ over the bandwidth $\nu = 40 - 200$~MHz, presented in \cite{Cohen} and used by \cite{SARAS_Const}. The models are provided by A. Fialkov and are publicly available at: \url{https://people.ast.cam.ac.uk/~afialkov/Collab.html}. We add to these models a foreground given by a $\nu^{-2.5}$ to produce simplistic mock data sets. The data sets are noiseless and while this is unrealistic, we would not expect the addition of noise to obscure any larger signal structure present in the data from a 21-cm experiment with a low radiometer noise. We fit each simulated data set with an MSF, low order unconstrained polynomials and a set of PSFs. All of the fitted DCF foreground models are 13\textsuperscript{th} order and of the form given by \cref{eq:loglog_poly}. We test all sign combinations for the DCF fits in this section for reasons that were explained in \cref{sec:badproblems} and find that the chosen DCF model and order provides the best fits after testing the built-in \maxsmooth~models.

\cref{fig:PSF_comp}, left panel, shows the difference in the number of turning points, $\Delta p$ for the residuals, $p_\mathrm{Fit~Residuals}$, and for the signal, $p_\mathrm{Signal}$, using four different foreground models as a function of the maximum brightness temperature, $T_\mathrm{max}$, and minimum temperature, $T_\mathrm{min}$, of the simulated signal. Each data point corresponds to one of the $264$ mock data sets and $\Delta p = 0$ signifies that the residuals have the same number of turning points as the signal. The unconstrained polynomial fits have the same functional form as the DCFs.

We can quantify the probability of returning residuals with the same number of turning points as the model signals. \cref{tab:table1} shows the total number of residuals for each fit type that returned the same $p$ as the simulated signals and 1 or 2 additional turning points. The MSF fits return $p_\mathrm{Signal}$ for $44\%$ of the mock data sets and $99\%$ of the time it returns at most $p_\mathrm{Signal}$ plus 2 additional turning points. They are the most likely, of the tested foreground models to return the correct number of turning points. The equivalent figures for the 5\textsuperscript{th} order logarithmic unconstrained polynomial, one of the most frequently used foreground fits in 21-cm cosmology, are $0\%$ and $64\%$ and for the PSF with $m \geq 3$ they are $0.004\%$ and $85\%$. The statistics suggest that modelling foregrounds with MSFs and well constrained PSFs can frequently result in residuals that closely follow the structure of the signal. We include in the statistics the cases with 1 or 2 additional turning points because the signals should still be identifiable in the residuals. A joint fit of foreground model plus a signal model in these instances should return an approximately correct parameterisation of the signal model.

\begin{table}
    \centering
    \begin{tabular}{ccccc}
        \hline
          & $m\geq2$ & $m\geq3$ & $m\geq4$ &$N = 5$ \\
          & (MSF) & (PSF) & (PSF) & (Poly) \\
         \hline
         $p_{\mathrm{~Signal}}$ & 116 & 1 & 0 & 0 \\
         $p_{\mathrm{~Signal}}+1$ & 132 & 135 & 15 & 52 \\
         $p_{\mathrm{~Signal}}+2$ & 14 & 88 & 147 & 116 \\
    \end{tabular}
    \caption{The table shows the total number of fits, using one of four foreground models, to the smooth foreground plus signal simulations from \protect\cite{Cohen} that have residuals with the same number of turning points as the signal, $p_\mathrm{~Signal}$. The data corresponds to that shown in the left panel of \cref{fig:PSF_comp}. We also show instances where the recovered residuals have one or two additional turning points.}
    \label{tab:table1}
\end{table}

\subsubsection{Example Residuals}

Also shown in \cref{fig:PSF_comp}, right panel, is an example of the fits for a given model and this is akin to Fig.7 in \cite{MSFCD}. We see that an MSF, top panel, while not identically recovering the signal but rather a smooth baseline subtracted version does preserves the three turning points of the model signal as expected. The example residuals from unconstrained polynomial fits shown in the bottom right panel of \cref{fig:PSF_comp} show a larger disparity with the structure of the model 21-cm signal than the MSF fit. The PSF with derivatives of order $m \geq 3$ constrained,  middle panel, produces residuals with one additional turning point. The behaviour at low frequency of the DCF fits is consistent across all of the 264 tested models. It is a byproduct of the basis choice, the frequency of data sampling and the steep nature of the foreground at low frequencies. We can alleviate some of these issues by increasing the sampling rate of our mock experiment and by reducing the bandwidth. However, the dominant cause is the basis function choice.

In logarithmic space any non-smooth variations in the data and derivatives are amplified as shown in \cref{sec:badproblems}. Since the mock data set here is noiseless, the only non-smooth structure comes from the signal. However, the data is predominately smooth at high frequency and so the optimum fit tends to be an accurate representation of the foreground in this region and poorer at lower frequencies. The additional turning points when comparing the MSF and PSFs are also seen at low frequencies for the same reason. Despite the above we maintain the full bandwidth, the same sampling rate and the logarithmic basis function in this analysis. We do this because this basis function gives us the best fitting DCF models and in a real experiment our knowledge of any present signal of interest will be too poor to constrain the bandwidth.

We have not analysed the full family of possible PSFs or unconstrained polynomials. Removing constraints on the lower order derivatives first will have the largest affect on the structure of the residuals and consequently we have extensively analysed the effects of lifting the restrictions on the 2\textsuperscript{nd} and 3\textsuperscript{rd} order derivatives. We have, however, found evidence to suggest that MSFs and well constrained PSFs can recover signal structure to a higher degree of accuracy than unconstrained polynomials in the case of a smooth foreground. If the foreground features additional non-smooth structure we may expect that an appropriately constrained PSF will act as an MSF. Determination of the appropriate constraints on a PSF will depend on the structure of the data, the expected structure of the signal in the data and the quality of the fit in terms of $\chi^2$.

\subsubsection{Identifiable 21-cm Signals and Limitations of DCFs}
\label{sec:limitations}

The left panel of \cref{fig:PSF_comp} is useful for 21-cm cosmology, in characterising the signals most likely to be detectable using DCFs. We find that MSFs and $m \geq 3$ PSFs will best recover signals with approximately $T_\mathrm{min} \geq -225$~mK and $T_\mathrm{max} \geq 0$~mK. We would expect this to be true generally because these signal models have complex structures and feature the strongest deviations from the smooth foreground. For the coldest models with $T_\mathrm{min} \leq -225$~mK and $T_\mathrm{max} \leq 0$~mK, X-ray heating is negligible and the spin temperature is always seen in absorption against the CMB. Consequently, they have the simplest structure, a weak deviation from the smooth foreground and are likely to be fitted out as part of the foreground modelling.

Comparison to restrictions placed on the most probable structure of the Global 21-cm signal from experimental data will help identify whether DCFs can recover these signals. For example \cite{SARAS_Const} ruled out models with low X-ray heating and $T_\mathrm{max} < 0$~mK, this is supported by results presented in \cite{EDGES-HB}. This suggests that DCFs are well suited to identify the most probable 21-cm signals. However, X-ray heating is only one of the structure defining processes. A more thorough exploration of this in terms of signal model parameters, such as star-formation efficiency and X-ray luminosity, is needed to fully understand the types of theoretical 21-cm signals that DCFs are sensitive to. This is out of the scope of this paper and will be the subject of future work.

Smooth systematics, like smooth 21-cm signals, in the data will be removed or fitted out as part of the foreground. However, unless independent modelling of systematics is required, this can be considered an advantage. Modelling foregrounds with DCFs will help to identify non-smooth systematics in data sets where unconstrained polynomials have the potential to fit these out. This is particularly important in 21-cm cosmology where these types of systematics need to be identified and instrumentation needs to be iteratively improved to increase the chances of a detection.

\subsubsection{Smooth Signal Models}

\begin{figure*}
\centering
    \includegraphics{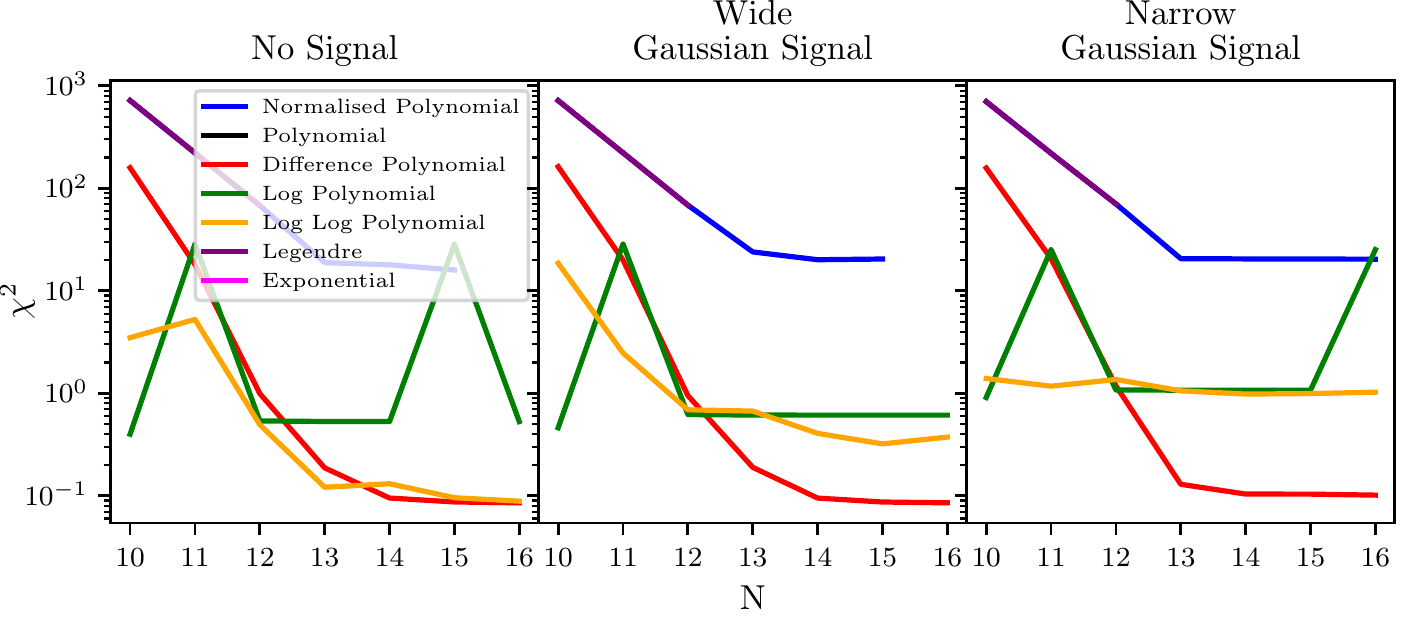}
    \caption{A comparison of the quality of MSF fit produced using the sign navigating algorithm and the different built-in \maxsmooth~basis as a function of order $N$ for three mock 21-cm experiment data sets with no signal, a wide Gaussian signal and a narrow Gaussian signal. The Difference Polynomial model, \cref{eq:additional_basis_b} of order $N = 15$ is shown to be the optimal model for fitting these data sets. Beyond $N = 15$ the quality of fit does not improve any further and additional terms in the model have coefficients $a_{k \geq 14} \approx 0$.}
    \label{fig:fig9_bestbasis}
\end{figure*}

\begin{figure*}
    \centering
    \includegraphics{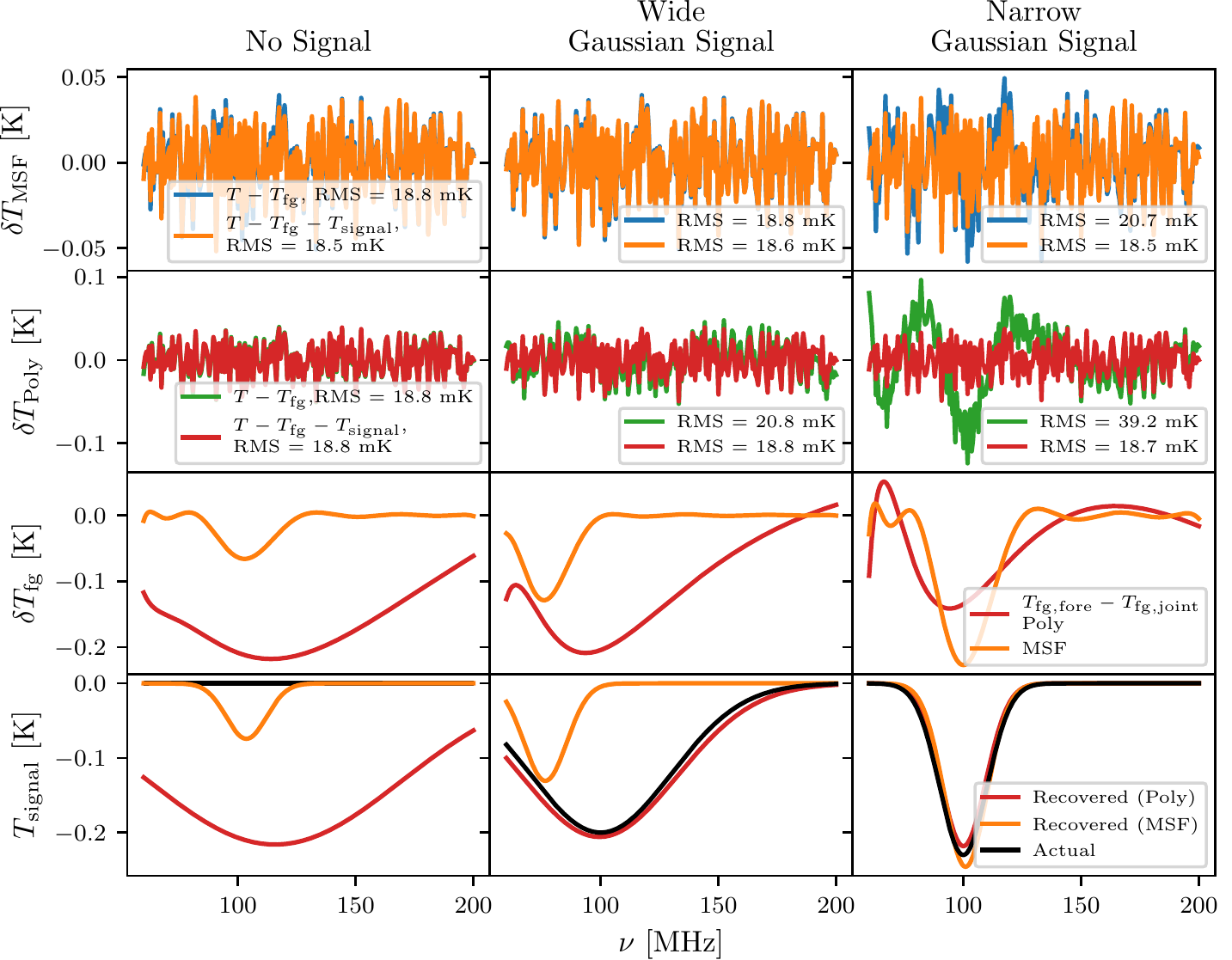}
    \caption{The top row shows the resultant residuals when using an MSF to fit just a foreground and to jointly fit a foreground and Gaussian signal model to three mock data sets with no signal, a wide Gaussian signal and a narrow Gaussian signal from left to right. The second row shows the equivalent for an unconstrained polynomial fit. The third row shows the change in foreground, $\delta T_\mathrm{fg}$, between the pure foreground fit, $T_\mathrm{fg, fore}$, and joint fit, $T_\mathrm{fg, joint}$, for both foreground models and all three signal models. The bottom row shows the recovered signals for both foreground models alongside the true signal models in the data sets. The figure appears to show that unconstrained polynomials are better behaved than MSFs, however we would note that the signal models used here are simplistic in nature. We would expect the signals to be more complex, including emission, as with those used in \cref{fig:PSF_comp} for which we showed the use of DCFs to be advantageous.}
    \label{fig:smooth_gauss}
\end{figure*}

For 21-cm cosmology it is also important to consider how the Global 21-cm signal is modelled when performing joint fits. Typically the signal is modelled as a Gaussian, flattened Gaussian or using physically motivated models. If a Gaussian model is jointly fit with a DCF foreground, then the fit is biased towards returning a `smooth' Gaussian signal with a large variance, $\sigma$, or full width at half max, $\mathrm{FWHM} = 2\sqrt{2\log(2)}~\sigma$, even if such a signal is not real. Similarly, incorrectly constrained DCFs can fit out `smooth' signals in data sets. This can cause uncertainty in the presence of such signals and the point is furthered in \cref{sec:EDGES_fits}.

We can illustrate this by generating three different data sets all with foregrounds following $\nu^{-2.5}$ and the same Gaussian noise with a standard deviation of $20$~mK. Into two of the three data sets we add mock Gaussian signals with central frequencies $\nu_c = 100$~MHz and generated using
\begin{equation}
    T_{21} = -A \exp\bigg(-\frac{(\nu - \nu_c)^2}{2\sigma^2}\bigg),
\end{equation}
where $A$ is the amplitude. The first mock signal has an amplitude of $200$~mK and a variance of $30$~MHz representing a realistic wide or smooth Gaussian 21-cm signal. The second represents a narrow Gaussian signal with an amplitude of $230$~mK and a variance of $10$~MHz. The final data set has no additional signal in the band of $60 - 200$~MHz.

\begin{table*}
    \centering
    \begin{tabular}{cccccc}
        \hline
          Model& $(\delta T_\mathrm{fg} - T_\mathrm{sig})_\mathrm{max}$ & $\Delta RMS$ & $\log(Z_\mathrm{fore})$ & $\log(Z_\mathrm{joint})$ & $\Delta \log(Z)$\\
          & [mK] & [mK] \\
         \hline
         \multicolumn{6}{c}{No Signal}\\
         \hline
         MSF & 6.6 &  0.2 & 606.42 $\pm$ 0.03 & 606.85 $\pm$ 0.03 & 0.08 $\pm$ 0.04\\
         Poly & 4.5 &  0.1 & 604.28 $\pm$ 0.03 & 603.38 $\pm$ 0.03 & 0.51 $\pm$ 0.04\\
         \hline
         \multicolumn{6}{c}{Wide Gaussian Signal}\\
         \hline
         MSF & 6.0 &  0.2 & 606.46 $\pm$ 0.03 & 606.57 $\pm$ 0.03 & 0.17 $\pm$ 0.04\\
         Poly & 17.5 &  2.6 & 575.02 $\pm$ 0.03 & 597.85 $\pm$ 0.05 & 22.98 $\pm$ 0.05\\
         \hline
         \multicolumn{6}{c}{Narrow Gaussian Signal}\\
         \hline
         MSF & 18.4 & 1.7 & 588.40 $\pm$ 0.03 & 593.42 $\pm$ 0.05 & 7.44 $\pm$ 0.06\\
         Poly & 81.4 & 20.3 & 432.69 $\pm$ 0.03 & 589.88 $\pm$ 0.06 & 157.10 $\pm$ 0.07\\
    \end{tabular}
    \caption{The table shows the average, across 10 repeats with different random noise, maximum difference between the recovered signals and the change in foreground for the polynomial and MSF pure foreground and joint fits, $(\delta T_\mathrm{fg} - T_\mathrm{sig})_\mathrm{max}$, to the three simulated 21-cm experiment data sets with no signal, a wide signal and a narrow Gaussian signal. Also shown are the average changes in RMS, $\Delta RMS$, between the pure foreground and joint fits for both foreground models and all three data sets. We also provide weighted average log-evidences and values for the change in log-evidence, $\Delta \log(Z)$, between the joint and pure foreground fits with errors calculated from the error in each recovered $\log(Z)$. The comparatively large change in log-evidence for the narrow Gaussian signal data set when using MSFs provides confidence that the recovered signal is truly present in the data. Unconstrained polynomials appear to perform better than MSFs in this analysis. However, the signals used here are simplistic and we have shown in \cref{fig:PSF_comp} that DCFs behave far better for complex theoretically motivated signals that include emission.}
    \label{tab:gauss_summary}
\end{table*}

We assess the best basis from the \maxsmooth~library for fitting these data sets and this is shown in \cref{fig:fig9_bestbasis}. The graph shows the minimum $\chi^2$ values when fitting MSFs of a particular form, detailed in the legend, using the \maxsmooth~sign navigating algorithm with a given order $N$ to the three data sets. We find that for all three data sets the best basis choice is the Difference polynomial, \cref{eq:additional_basis_b}, with order $N = 15$. We consequently proceed to fit each data set with this MSF model. We compare the resultant residuals for each data set to those from a joint fit of an MSF, of the same functional form, and a Gaussian 21-cm signal model. We perform the joint fits by using \maxsmooth~with the Python implementation of the nested sampling software \multinest~\citep[][]{PyMultiNest,MultiNest,MultiNest2}. Here, \multinest~estimates the Gaussian signal parameters, \maxsmooth~fits the foreground model to the data minus the estimated Gaussian signal at each iteration and \multinest~minimises the data minus the foreground model plus the signal model. We also perform the equivalent fits with a 5\textsuperscript{th} order unconstrained polynomial given by \cref{eq:loglog_poly} using a Lavenberg-Marquardt \citep{Levenberg1944, Marquardt1963} algorithm in place of \maxsmooth. We perform this analysis ten times generating a new noise distribution each time. An example result is shown in \cref{fig:smooth_gauss} where we have provided the same theoretically motivated priors on all of the Gaussian 21-cm models. Respectively from top to bottom the rows in \cref{fig:smooth_gauss} show the residuals after just an MSF fit and after a joint fit for comparison, the equivalent for the unconstrained polynomial fits, the change in foreground between the foreground fit and the joint fit and the recovered signal in comparison to the actual signal. The columns correspond to the case of no signal, a wide Gaussian signal and a narrow Gaussian signal from left to right.

We can see that in the absence of a signal jointly fitting with a Gaussian model and MSF returns an absorption trough. This recovered absorption trough is approximately the same as the change in the foreground model when fitting with just an MSF and the change in RMS is small. This is illustrated in \cref{tab:gauss_summary} which shows the average results across the ten repeats of this analysis. If this were a real experiment the fitted model could easily be misinterpreted as a real signal that had been fitted out as part of the foreground model. However, there is a very small change in log-evidence between the pure foreground and joint fit which would suggest that the signal is not truly present in the data.

In \cref{tab:gauss_summary} the values for the log-evidence for each fit and the change in log-evidence, $\Delta \log(Z)$, between the foreground and joint fits for each data set and each foreground model are reported. While \multinest~returns a log-evidence for the joint fits, \maxsmooth~is not a Bayesian algorithm. However, assuming a Gaussian likelihood we can use \multinest~to calculate the evidence for our pure foreground fits by having it estimate the noise and calculate the likelihood.

When jointly fitting in the presence of a wide Gaussian signal with an MSF we see a similar result to the case with no additional signal and we could not confidently say that the signal is present in the data if we had no prior knowledge. In fact, the joint fit has recovered a poor representation of the Gaussian signal because the smooth signal in both instances, pure foreground fit and joint, has been absorbed in the foreground modelling.

Finally, we see that in the case of a narrow Gaussian signal in the data set with an MSF we get an almost exact recovery after a joint fit. There is a larger discrepancy between the difference in foreground models and the recovered signal, as shown in \cref{tab:gauss_summary}, and the reduction in RMS is more significant giving us confidence that the signal is truly present. Importantly, the change in log-evidence is much larger than in the other two cases indicating the presence of the signal.

For the unconstrained polynomial, in the case where there is no signal in the data we appear to recover a very smooth Gaussian signal. However, again the change in log-evidence between the pure foreground fit and the joint fit tells us that neither scenario is more likely and consequently we would conclude that the signal is not present. For the other two cases the unconstrained polynomial behaves well. However, we note that the signals induced here are simplistic representations of a Global 21-cm signal. We would expect the signal to have a much more complicated structure including emission at high frequencies and similar to the signals used in \cref{fig:PSF_comp} for which we showed DCFs to be advantageous.

When attempting to identify the wide Gaussian signal using DCFs we also note it will be advantageous to fit using a CSF with derivatives of order $m \geq 1$ constrained. For 21-cm cosmology we find that CSFs and MSFs are generally equivalent. However, in this case the only significant non-smooth structure in the signal, aside from the inflection point, is the turning point and where an MSF may fit this out a CSF will not.

The problem of identifying and misidentifying smooth signals becomes even more prominent, particularly in the absence of any signal in the data, when modelling the foreground with a DCF in $T - \log(\nu)$ space. Here the sample rate is non-uniform and higher at higher $\log(\nu)$. This makes the problem harder to fit in this region and also `smooths' any signals present at low $\log(\nu)$. Together these effects can make it difficult to detect signals that can already be considered `smooth' in linear space. Similarly, in the absence of any signal in the data, when jointly fitting with a Gaussian signal model and large prior ranges, the routine estimating the Gaussian parameters will tend to favour `smooth' signals at high frequencies in linear space because of the non-uniformity of sampling in logarithmic space.

\subsection{MSFs and the EDGES Data}
\label{sec:EDGES_fits}

In 2018 the EDGES team reported the detection of an absorption trough at $78$~MHz which could be interpreted as a Global 21-cm signal \citep[]{EDGES_LB}. The reported signal is $\approx 2$ times the maximum magnitude predicted by current cosmological models \citep[]{Cohen}, and, in order to explain the signal as a 21-cm signal, interactions between dark matter and baryons \citep{BarakanaDM2018, BerlinDM2018, KovetzDM2018, MunozDM2018, SlatyerDM2018} or a higher radio background \citep{EDGES_LB, EwallRB2018, FengRB2018, JanaRB2018, FialkovRB2019, MirochaRB2019} are needed.

\begin{figure}
    \centering
    \includegraphics{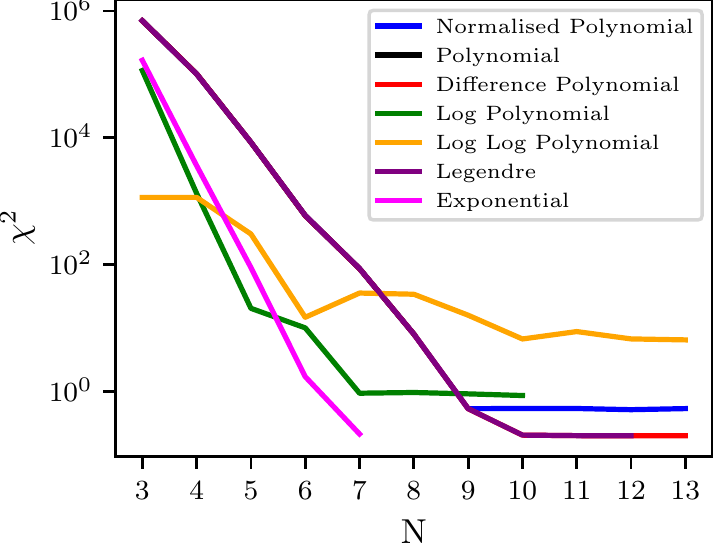}
    \caption{The resultant $\chi^2$ as a function of MSF order, $N$, for the \maxsmooth~built-in basis functions fitted to the EDGES data using the \maxsmooth~sign navigating algorithm. The Legendre, Difference polynomial, the Polynomial and Normalised Polynomial models lie on top of each other in this figure. The occasional increase in $\chi^2$ with $N$ for the logarithmic model is because this basis is increasingly unstable with higher $N$ and requires all sign combinations to be tested.}
    \label{fig:EDGES_basis}
\end{figure}

While a higher radio background has been suggested by the results of the ARCADE-2 experiment \citep{ARCADE2} and confirmed by measurements from LWA \citep{LWA_radio_background} there are concerns about the analysis of the EDGES data \citep{Hills, Sims, MSF-EDGES}. These studies and the following work presented here use the publicly available integrated spectrum from the EDGES Low Band experiment
which can be found at: \url{https://loco.lab.asu.edu/edges/edges-data-release/}.

\begin{figure*}
    \centering
    \includegraphics{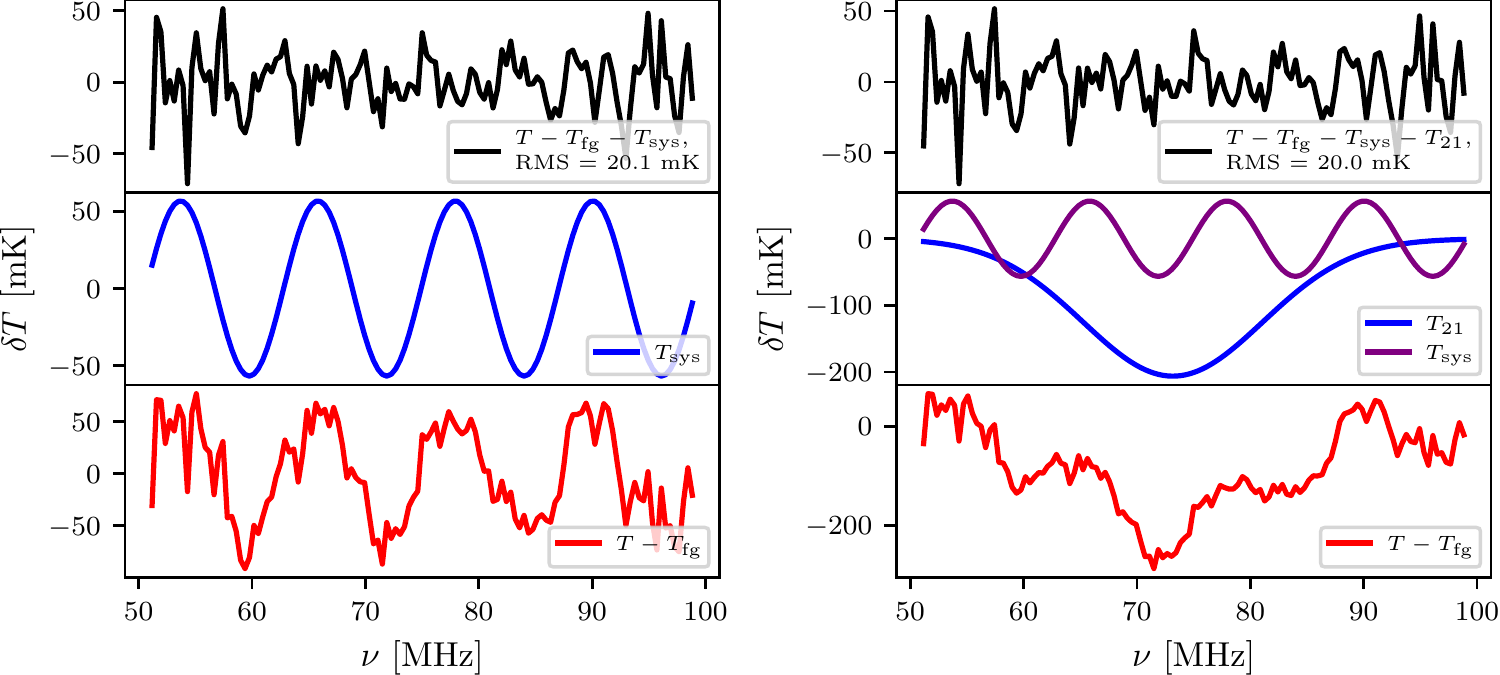}
    \caption{\textbf{Left:} The residuals with an RMS of $20.1$~mK and log-evidence of $302.99\pm0.08$, black, found after subtracting from the EDGES data a jointly fit MSF and a sinusoidal systematic. We recover a parameterisation of the systematic, shown in blue, that is consistent with previous work as discussed in the text. The red line, bottom panel, illustrates the residuals after just removing the fitted foreground from the data. \textbf{Right:} A joint fit of a Gaussian 21-cm model signal with a sinusoidal systematic and an MSF foreground to the EDGES data. The recovered 21-cm signal is smooth and the addition of the signal model has caused a large change in the recovered foreground, $T_\mathrm{fg}$, illustrated by the decreased amplitude in the bottom panel around $70$~MHz when compared to that in the left figure. The RMS, $20.0$~mK, is also similar in magnitude to that found for the fit in the left panel and the log-evidence has reduced to $292.67\pm0.17$. Consequently, it is unlikely the model signal is real and it is probable that it is produced artificially by the change in foreground.}
    \label{fig:EDGES_fits}
\end{figure*}

\cite{Hills} found that recovering the absorption profile using the `physically motivated' foreground model in \cite{EDGES_LB} produces unphysical negative values for the ionospheric electron temperature and optical depth. This suggests that the treatment of the foreground in \citeauthor{EDGES_LB} absorbs part of an unknown systematic. It was also found that a large change in foreground was needed when just fitting a `physical' foreground to the data and when jointly fitting with a flattened Gaussian 21-cm signal profile. In \citeauthor{Hills} the authors identify the potential presence of a sinusoidal function in the EDGES data with an amplitude of $\approx 60$~mK and a period of $\approx 12.5$~MHz.

\cite{Sims} fit a range of models to the EDGES data varying the 21-cm models between a Gaussian model, a flattened Gaussian model as used by \cite{EDGES_LB} and physical simulations from the ARES code \citep{ARES_sim}. They vary the unconstrained polynomial order for the foreground model and examine likelihoods with and without an additional noise term and a damped sinusoidal function. They use Bayesian Evidence to quantify the most likely scenarios of an atlas of 128 models. The 21 highest evidence models all feature damped sinusoidal functions all with a consistent amplitude of $\approx~60$~mK and a period of $\approx~12.5$~MHz.

An MSF fit to the foreground should leave a periodic sinusoidal function behind in the residuals if it is present in the data because it is non-smooth in nature. This has previously been shown to be the case by \cite{MSF-EDGES}, hereafter S19, who identified a sinusoidal feature with an amplitude of $60 \pm 10$~mK and a period of $12.3 \pm 0.1$~MHz. We attempt here to re-create this analysis to illustrate the abilities of \maxsmooth. The fitting routine used and choice of basis function are the only differences between the results presented here and in S19. The use of \maxsmooth~means that our joint fit of the data and a systematic model will be computationally quicker and more reliable than the Nelder-Mead based approach to fitting taken in S19, as demonstrated in \cref{sec:Eff}.

We begin first by assessing the quality of fits using the various basis functions built-in to \maxsmooth. S19 used a basis function constrained in $T - \log_{10}(\nu)$ space and although the functional form is not explicitly stated in S19 it is derived from the models in \cite{MSFCD} and so is similar to, if not identical to, \cref{eq:log_poly}. \cref{fig:EDGES_basis} shows the resultant $\chi^2$ as a function of MSF order for fits with varying basis functions using \maxsmooth. This figure again shows how the choice of basis function can affect the quality of the fit. Of note is that our $T - \log_{10}(\nu)$ space model cannot achieve the same RMS as that found by S19 with a similar model. With an $N = 7$ MSF constrained in logarithmic frequency space S19 return an RMS of $44$~mK, whereas with \cref{eq:log_poly} \maxsmooth~returns an RMS of $87$~mK. We believe this is due to the lack of normalisation in \maxsmooth~and as previously discussed this is an ongoing area of development.

We use \cref{fig:EDGES_basis} to inform our choice of basis function and MSF order and proceed using an 11\textsuperscript{th} order MSF of the form given by \cref{eq:additional_basis_b}. We find residuals with an RMS of $\approx~40.4$~mK and a log-evidence of $216.80\pm 0.09$ as shown in \cref{fig:fig0}. We note that this is in approximate agreement to results shown in S19. We also find troughs at $\approx~70$~MHz and $\approx~85$~MHz which correspond to those found in all of the reported sinusoidal functions.

We jointly fit the data with an 11\textsuperscript{th} order MSF and a sinusoidal function of the form
\begin{equation}
    T_\mathrm{sys}~=~p_0~\sin(p_1~\nu~-~p_2),
    \label{eq:sine_sys}
\end{equation}
and the resultant residuals are shown in the left panel of \cref{fig:EDGES_fits}. Note we have not included a model 21-cm signal in this fit. We use \maxsmooth~along with a Lavenberg-Marquardt algorithm implemented with \scipy~to perform this joint fit and with initial parameters of $p_0 = 60$~mK, $p_1\approx (2\pi)/12.5$~MHz$^{-1}$ and $p_2 = 0$~rad. We find that the results change with the initial parameters when using the Lavenberg-Marquardt algorithm, however, the chosen initial parameters are well informed by the previous work outlined above. We return parameters of $p_0 \approx 56.6$~mK, $p_1 \approx 0.52~\textnormal{MHz}^{-1}$ or a period of $ \approx 12.1$~MHz and $p_2 \approx 1.1$~rad in close agreement with previous analysis. We use \multinest~to approximate the evidence for this fit assuming a Gaussian likelihood and return $\log(Z)=302.99\pm0.08$. This is a significant increase in log-evidence when compared to the pure foreground fit and would suggest strongly that the systematic is present in the data.

We find an RMS value of $20.1$~mK, in close agreement with the result of $22.9$~mK found in S19 when jointly fitting an MSF and the sinusoidal function. However, we note that the RMS of the joint fit in the left panel of \cref{fig:EDGES_fits} is equivalent to the RMS found by S19 when jointly fitting an MSF foreground, a sinusoidal systematic and Gaussian 21-cm signal model. Their proposed Gaussian 21-cm signal model fits with standard predictions. However, the RMS of our joint fit without a Gaussian signal model may highlight some of the difficulties in detecting `smooth' Gaussian signals discussed in \cref{sec:limitations}.

We perform a joint fit of a Gaussian 21-cm signal, a sinusoidal systematic and MSF. Due to the increased complexity of the fit and uncertainty in the fit parameters for the Gaussian model we perform our fit using \maxsmooth~and \multinest. We provide prior ranges of $30 - 80$~mK for the amplitude, $0 - 25$~MHz for the period and $0 - 2\pi$ for the phase shift of the sinusoidal systematic. For our Gaussian we set realistic priors on the amplitude of $0 - 250$~mK, on the central frequency of $50 - 100$~MHz and on $\sigma$ of $0 - 20$~MHz. For the sinusoid we find an amplitude of $56.5$~mK, a period of approximately $12.1$~MHz and a phase of $1.1$~rad. These results, using a more extensive search of the parameter space, are consistent with our previously found sinusoid further indicating that when performing the fit shown in the left panel of \cref{fig:EDGES_fits} our initial parameters were well chosen. We return an amplitude of $206$~mK for the Gaussian with a central frequency of $73$~MHz and a $\mathrm{FWHM} \approx 18$~MHz. \multinest~returns a noise parameter of $20$~mK and our resultant fit, shown in the right panel of \cref{fig:EDGES_fits}, has an RMS of approximately $20.0$~mK. This is much wider and deeper than the signal reported in \cite{MSF-EDGES} which has a depth of $133 \pm 60$~mK and FWHM of $9 \pm 3$~MHz however we return the same central frequency of $73$~MHz.

Noting the discussion of plausibly detectable 21-cm signals when using DCFs and the bias towards `detection' of `smooth' Gaussian signals in \cref{sec:limitations} we assess the feasibility of the returned model signal. Here the notion of `smoothness' is relative to the bandwidth. In \cref{sec:limitations} a Gaussian with $\mathrm{FWHM} \approx 24$~MHz is confidently identifiable as a real signal but the bandwidth is much larger than that for the EDGES data. We can see by comparison of the bottom panels of \cref{fig:EDGES_fits}, showing the data minus the foreground from the joint fits, that there is a large change in foreground when we include the Gaussian model as part of our joint fit. This may be because in our initial fit, without the Gaussian model, the foreground model was fitting out the smooth signal. Alternatively, the signal may not be present and the fitting routine has returned a smooth signal by extracting it from the foreground component of the fit. The reduction in RMS when the joint fit includes a Gaussian is $0.054$~mK and it is consequently challenging to determine whether or not the signal is present in the data. However, we can conclude from the log-evidence, which for this fit has a value of $292.67\pm0.17$ and is smaller than that for the joint fit of the systematic and foreground, that the Gaussian is not likely to be real.

\subsection{MSFs and the LEDA Data}

LEDA, like EDGES, is a radiometer based Global 21-cm experiment analysing the band $30 - 88$~MHz and aiming to detect the anticipated absorption feature \cite{Greenhill2012}. The design and calibration approach of LEDA is detailed in \cite{LEDA}. In contrast to EDGES, the LEDA experiment is comprised of 5 dual-polarization radiometer antennas that are part of a larger 256-antenna interferometric array. This approach is intended to allow inter-antenna comparison and in-situ measurement of the antenna gain response. Similar to other radiometry experiments with absolute calibration, LEDA uses two noise diode references to calibrate the measured antenna temperature into units of Kelvin. Corrections are then applied to account for the impedance of the antenna and receiver, derived from vector network analyser~(VNA) measurements. 

As shown in \cite{LEDA}, data are seen to vary between antennas, which are not perfectly identical, and this is attributed to minor differences in terrain, analog component response, and physical construction. While calibrated spectra are presented, it is suggested that there are unidentified systematics in the data; work has been undertaken to better characterize and update the LEDA system with iterative improvements. Further measurements were taken in 2017 and 2018, which are under analysis (Gardsen et al., in prep.). 

Here, we fit MSFs to data from the LEDA 2016 campaign \citep[][Spinelli et al., in prep.]{LEDA}. This is the first time MSFs have been applied to LEDA data. Specifically, we fit data from antenna 252A, taken on January 26\textsuperscript{th} 2016 in the LST range 11:00-12:00.  In this LST range, the Galactic contribution to the antenna temperature is at a minimum. The data are binned into 1.008 MHz channels, spanning 40--85 MHz.

\begin{figure}
    \centering
    \includegraphics{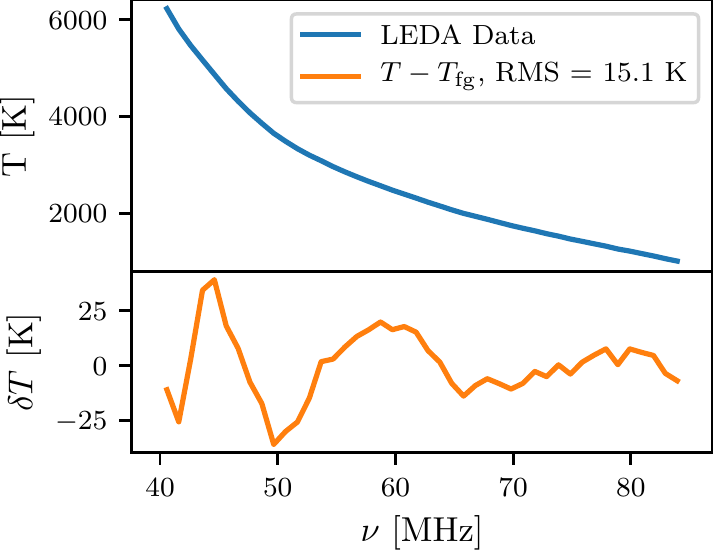}
    \caption{The LEDA data from antenna 252A taken in 2016 and averaged over 1 hour of LST shown in the top panel. Also shown, bottom panel, are the residuals after fitting the LEDA data with a 9\textsuperscript{th} order MSF of the form given in \cref{eq:additional_basis_b} with a log-evidence of $-185.45\pm0.09$. We see evidence of a damped sinusoidal systematic in the data set.}
    \label{fig:LEDA_data}
\end{figure}

We fit an MSF of the form given in \cref{eq:additional_basis_b} to the data as we find that this basis function returns the best fit consistently for $N \geq 8$. Shown in \cref{fig:LEDA_data} are the resultant residuals from an $N = 9$ MSF fit with an RMS of $\approx 15$~K and a log-evidence of $-185.45\pm0.09$. The resultant residuals are large and would obscure a cosmological 21-cm signal. We note that as per equation~(4) in \cite{LEDA} the radiometer noise is expected to be $\approx 0.5$~K.

\begin{figure*}
    \centering
    \includegraphics{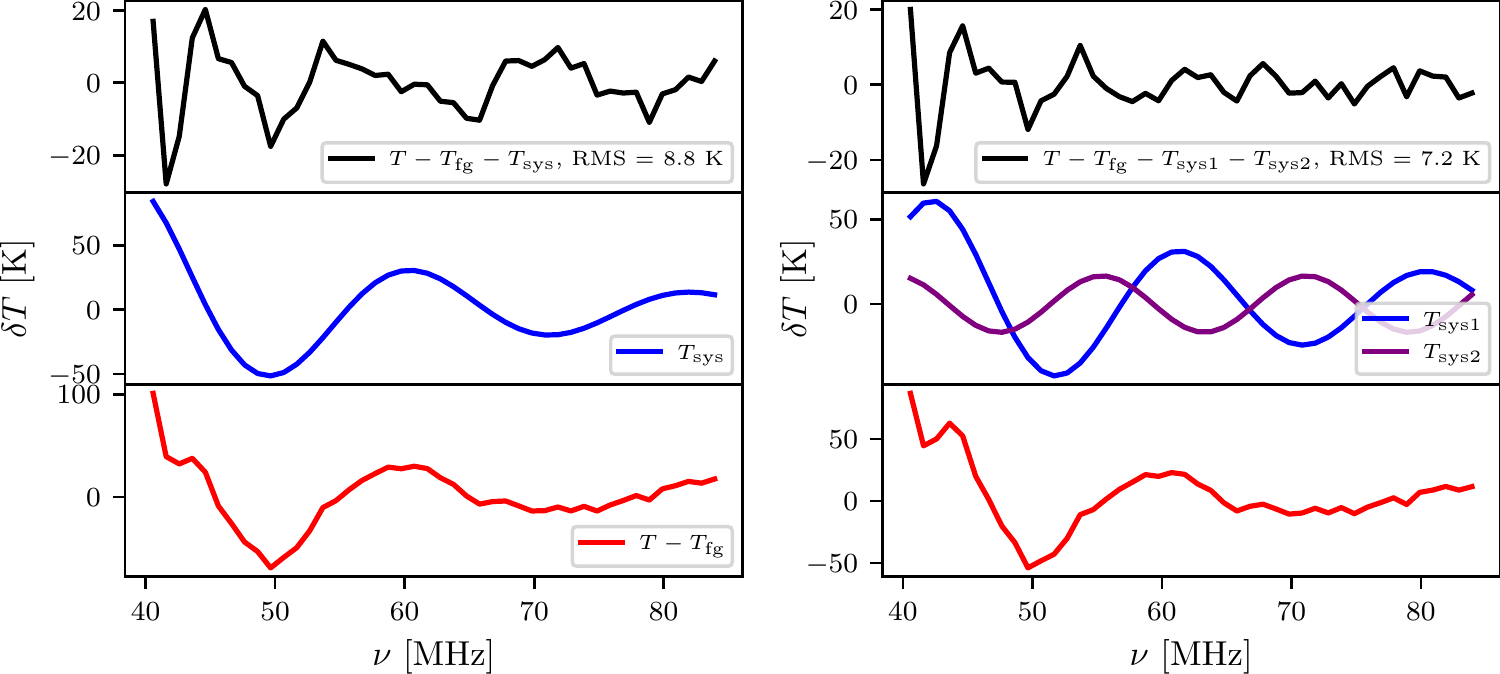}
    \caption{\textbf{Left:} The residuals, black, after jointly fitting the LEDA data with an MSF and damped sinusoidal systematic with a log-evidence of $-175.50\pm0.19$. The centre panel shows the recovered systematic model, blue, and the bottom panel shows the residuals, red, after just subtracting the fitted foreground model. The addition of the systematic model has reduced the RMS of the fit when compared to \cref{fig:LEDA_data}. \textbf{Right:} The resultant residuals, black, with a log-evidence of $-168\pm0.19$ found when fitting the LEDA data with an MSF foreground, damped sinusoidal, blue, and additional sinusoidal systematic, purple. Again the bottom panel shows the residuals, red, after just subtracting the fitted foreground model from the data. The further reduction in RMS and increase in log-evidence suggests that both these systematics are present in the data and indicates that the larger systematics in the LEDA data may be represented by the leading order terms in a damped Fourier series.}
    \label{fig:LEDA_resid}
\end{figure*}

The residuals from the MSF fit clearly feature a damped sinusoidal systematic. We proceed to fit a systematic model given by
\begin{equation}
    T_\mathrm{sys}~=~\bigg(\frac{\nu}{\nu_0}\bigg)^{-p_0} p_1~\sin(p_2~\nu~-~p_3),
    \label{eq:damp_sine_sys}
\end{equation}
along with a 9\textsuperscript{th} order MSF by using \maxsmooth~and \multinest. $\nu_0$ is chosen to be the central frequency of the band. We provide a prior on the power of $0 - 3$ for weak damping. Prior ranges of $25~-~75$~K for the amplitude of the sinusoidal function, $0~-~1$~MHz$^{-1}$ for the period, $P$, which is fitted as $p_2=(2\pi)/P$, $0~-~2\pi$ for the phase shift and a log uniform prior on the noise of $10^{-2}~-~10^{1}$~K are also provided. The results of this fit are shown in the left panel of \cref{fig:LEDA_resid}. We return optimal parameters of a exponent of $\approx 2.7$, an amplitude of $\approx 27.9$~K, a period of $21.7$~MHz and a phase shift of $3.7$~rad. The residuals have an RMS of $\approx 8.8$~K and \multinest~returns a noise parameter of $\approx 7.7$~K. We also see an increase in log-evidence, which has a value of $-175.50\pm0.19$, when compared to the pure foreground fit suggesting the systematic is present in the data.

\cite{LEDA} suggest, from analysis of the 2016 LEDA data, that the systematic in the data is caused by the direction-dependent gain of the antenna. A frequency dependent group delay may also be caused by a bandpass filter that could contribute unaccounted for reflections. The pattern of oscillations that form the systematic have also been found to change after rainfall. This systematic may then be caused by moisture in the surrounding soil or by changes in the electric length of the dipoles caused by moisture on the dipole itself. We also highlight the similarities in structure of the systematic with that in the EDGES data. Both have sinusoidal structures and so similarities between the experimental setups and calibration processes may hint at larger causes of systematics across 21-cm cosmology experiments. The systematic is not likely to be associated with the sky because of the difference in periodicity and amplitude found by both experiments. EDGES does not have a bandpass filter, it could still be affected by moisture in the surrounding environment however we note that this experiment is in a typically dry location~(MRAO, Australia). 

The residuals shown in the top left panel of \cref{fig:LEDA_resid} show a further sinusoidal structure after removal of the leading order damped sinusoidal systematic. We therefore attempt a joint fit to the data using an MSF foreground, a damped sinusoid and an additional sinusoid described by \cref{eq:sine_sys}. We maintain the same priors on the original damped sinusoidal function and provide a prior of $10-30$~K on the amplitude, $0-1$~MHz$^{-1}$ on the period and $0-2\pi$ on the phase shift of the additional sinusoidal systematic.

We find best fit parameters for the leading damped sinusoid of $\approx 1.8$ for the exponent, $\approx 30$~K for the amplitude, a period of $\approx 19$~MHz and a phase shift of $\approx 6.2$~rad. For the additional sinusoidal systematic we find an amplitude of $\approx 17$~K, a period of $\approx 16$~MHz and a phase shift of $\approx 1.5$~rad. \multinest~returns a noise of $\approx 7.2$~K and the fit shown in the right panel of \cref{fig:LEDA_resid} has an RMS of $\approx 7.2$~K. We find a log-evidence for this fit of $-168.34\pm0.19$.

Distinctions between the two systematics have been made in the middle right panel of \cref{fig:LEDA_resid} for clarity. The RMS of the residuals after removal of these two systematics is still significantly larger than the radiometer noise for this experiment, $\approx 0.5$~K. However, the decrease in the RMS when these systematics are included in the fit and increase in log-evidence would strongly suggest that both are present in the data. A further addition of sinusoidal systematics will inevitably reduce the RMS of the residuals in the same way that the residuals after foreground removal could accurately be described by a Fourier series. Higher order terms in the series would feature smaller periods until the periodicity of the terms matched that of the noise. However, the systematics present in the data may have a form described by the leading order terms in a damped Fourier series as found here. We leave more rigorous investigation of the additional oscillatory structure in the residuals, top right panel of \cref{fig:LEDA_resid}, to future work.

\section{Conclusions}
\label{sec:conclusions}

Derivative Constrained Functions~(DCFs) generally are advantageous for experiments in which the desired signal is masked by higher magnitude smooth signals or foregrounds. A `smooth' foreground is one that follows a power law structure and DCFs are designed to accurately replicate this by constraining individual high order derivatives to be entirely negative or positive across the band of interest. They are particularly useful when the signal of interest is expected to be several orders of magnitude smaller than the foregrounds, similar in magnitude to the experimental noise and non-smooth in structure~(i.e., having high order derivatives that cross zero in the band of interest).

We have introduced \maxsmooth~as a fast and robust tool for fitting DCFs and demonstrated its abilities with examples from 21-cm cosmology. \maxsmooth~features a library of example DCF models which is designed to be extended. Further work into the normalisation of DCF models for \maxsmooth~is required with the aim to improve the quality of fitting and efficiency of the software.

In \cite{MSFCD} the authors fit Maximally Smooth Functions~(MSFs) using a Nelder-Mead routine to simulated sky data with Global 21-cm signals. They demonstrate that their fitting routine recovers the same residuals for 7\textsuperscript{th}, 10\textsuperscript{th} and 20\textsuperscript{th} order MSFs. \maxsmooth~is shown, however, to be capable of producing good fits $\approx2$ orders of magnitude faster than a Basin-hopping/Nelder-Mead based algorithm. This is an important improvement when jointly fitting signals, systematics and foregrounds using a Bayesian likelihood loop as in nested sampling \citep{Anstey2020}.

\maxsmooth~is also designed to be able to cover the entire available parameter space, unlike a Basin-hopping/Nelder-Mead based routine, by dividing it into discrete parameter spaces based on the different allowed combinations of signs, positive and negative, on the constrained derivatives. The extensive exploration of the parameter space provides confidence in the results and the employment of quadratic programming, a robust method for solving constrained optimisation problems, allows \maxsmooth~to remain an efficient algorithm.

We have reproduced analysis of the EDGES data using \maxsmooth~and analysed data from the LEDA experiment with MSFs for the first time. We have highlighted limitations of DCFs when jointly fitting for 21-cm signals and illustrated this using the EDGES data. We have shown that in the presence of a smooth signal or no signal that DCFs can incorrectly recover signals that are smooth across the band when jointly fitted with signal models. However, this is not a problem that is unique to DCFs and we have illustrated that it is of equal prevalence when using unconstrained polynomials. 

We show, also, that MSFs preserve turning points of 21-cm signals more consistently than commonly used low order logarithmic unconstrained polynomial models. This is particularly true of 21-cm signals with maximum brightness temperatures, $T_\mathrm{max} \geq 0$~mK and minimum temperatures, $T_\mathrm{min} \geq -225$~mK which feature the strongest deviations, a distinct absorption trough and emission above the background CMB, from the smooth foreground approximated by a $\nu^{-2.5}$ power law. A more detailed exploration of the signal parameter space is needed to fully understand the types of `detectable' or reproducible 21-cm signals when using DCFs with varying constraints to model the foreground.

Through the EDGES data and LEDA data we have illustrated that MSFs are useful in identifying non-smooth and periodic experimental systematics. This is advantageous for two reasons: it allows for better identification of any Global 21-cm signal present in the data and it allows the causes of the systematics to be better identified leading to iterative improvements in experimental setups. Where systematics with a smooth structure across the bandwidth of interest are also present we expect that these will be fitted out by DCF foreground models.

In the LEDA data, we have identified the presence of a damped sinusoidal systematic and additional sinusoid. We suggest that the similarities between the structure of systematics in the EDGES data and the LEDA data could highlight a larger issue in 21-cm experimentation. Further work is needed to identify a probable cause for such systematics and exploration of similarities between the approaches of the two experiments could help identify these causes with DCFs being the primary tool for foreground modelling.

We suggest here that DCFs may also be used as a tool for identifying low level RFI, weak spectral lines and as illustrated for MSFs by \cite{MSFRE}, signals from the Epoch of Recombination. In all cases the signals of interest are non-smooth features masked by higher magnitude smooth signals that can be modelled and removed with DCFs. Applications of DCFs in these fields is left for future work.

\section*{Acknowledgements}

HB thanks Peter Sims for useful discussions on the EDGES data and the nature of MSFs in different parameter spaces. HB would also like to thank Richard Hills and Annelies Mortier for discussion on section 5.1.4 and acknowledges the support of the Science and Technology Facilities Council~(STFC) through grant number ST/T505997/1. WH is supported by a Gonville \& Caius Research Fellowship. AF is supported by the Royal Society University Research Fellowship. EA is supported by the STFC through the SKA grant G100521.

\section*{Data Availability}

The EDGES data is available at \url{https://loco.lab.asu.edu/edges/edges-data-release/} and the simulated Global 21-cm signals used in this article are available at \url{https://people.ast.cam.ac.uk/~afialkov/Collab.html}. The LEDA data was provided by the LEDA collaboration.



\bibliographystyle{mnras}
\bibliography{References}

\appendix

\section{CVXOPT and Quadratic Programming}
\label{app:qp}

Quadratic programs are a special family of convex optimisation problem in which the objective function is quadratic and the conditions are affine in nature \citep[][]{cvx, qp}. \cvxopt~is a Python package for solving a quadratic optimisation problem subject to linear constraints. In \cref{sec:qp} we write the least-squares problem that we are solving in terms of matrices as
\begin{equation}
    \chi^2(\mathbf{a})~=~\frac{1}{2}~\mathbf{a}^T~\mathbf{Q}~\mathbf{a}~+~\mathbf{q}^T~\mathbf{a},
\end{equation}
where
\begin{equation}
    \mathbf{Q}~=~ \mathbf{\Phi}^T~\mathbf{\Phi}~~\textnormal{and}~~ \mathbf{q}^T~=~-\mathbf{y}^T~\mathbf{\Phi},
\end{equation}
subject to a constraint
\begin{equation}
    \mathbf{G}~\mathbf{a}~\le~\mathbf{h}.
    \label{eq:qp_cond}
\end{equation}
This is known as the primal problem when using quadratic programming to solve least-squares. For the constraints on a DCF $\mathbf{h} = \mathbf{0}$.

The problem is solved using the Karush-Kuhn-Tucker~(KKT) theorem \citep[]{KT1951, Karush2014} which re-phrases the above problem in terms of a Lagrangian given in this instance by
\begin{equation}
    \mathbf{L(a,\boldsymbol{\mu})}~=~\chi^2(\mathbf{a})~+~\boldsymbol{\mu^T}\mathbf{g(a)},
\end{equation}
where $\mathbf{g(a)}~=~\mathbf{G~a}-\mathbf{h}$ and $\boldsymbol{\mu}$ is the Lagrange multiplier.

From the Lagrangian we can define the Lagrangian dual function to be
\begin{equation}
    l(\boldsymbol{\mu}) = \min_{x} \mathbf{L(a,\boldsymbol{\mu})},
\end{equation}
which leads to the dual problem minimizing $l(\boldsymbol{\mu})$ subject to $\boldsymbol{\mu} \geq 0$.

The condition on the dual problem that $\boldsymbol{\mu} \geq 0$ is derived from the definition of the condition on the primal problem and the definition of $\mathbf{g(a)}$. The condition is known as complementary slackness and is given by
\begin{equation}
    \boldsymbol{\mu}\mathbf{g}(\mathbf{a})~=~0.
\end{equation}
Since $\mathbf{g}(\mathbf{a})~\le~0$, by definition this implies
\begin{equation}
    \boldsymbol{\mu}~\ge~0.
\end{equation}

The theorem states that if the point given by $(\mathbf{a^*}, \boldsymbol{\mu^*})$ is a saddle point in the Lagrangian in the domain with $\boldsymbol{\mu}~\ge~0$ then $\mathbf{a^*}$ is a solution to the optimisation problem. This is known as strong duality and can be re-phrased as
\begin{equation}
    \chi^2(\mathbf{a^*}) = l(\boldsymbol{\mu^*}).
\end{equation}

By taking the gradient of the Lagrangian and setting this equal to zero, since we are looking for a stationary point, we find
\begin{equation}
    \nabla \chi^2(\mathbf{a^*}) -\sum_i\mu_i^*\nabla g_i(\mathbf{a^*})~=~0,
    \label{eq:lagange_derive}
\end{equation}
where the sum is over the total number of different constraints. An optimal solution of the primal problem will be a stationary point with $\nabla \chi^2(\mathbf{a^*}) = 0$ and consequently we have $\sum_i\mu_i^*\nabla g_i(\mathbf{a^*})~=~0$ by \cref{eq:lagange_derive} leading to the required saddle point.

The algorithm consequently looks for solutions $\mathbf{a^*}$ for which a non-negative $\boldsymbol{\mu^*}$ can be found and the KKT conditions can be satisfied. To summaries the conditions are as follows,
\begin{enumerate}
    \item Stationary Condition: The optimal solution of the prime and dual problems will produce a saddle point in the Lagrangian.
    \item Complementary slackness: $\boldsymbol{\mu^*}\mathbf{g}(\mathbf{a^*})~=~0$  holds.
    \item Primal Feasibility: The condition given by \cref{eq:qp_cond} is satisfied by $\mathbf{a^*}$.
    \item Dual Feasibility: The Lagrangian multiplier satisfies the inequality $\boldsymbol{\mu^*}\geq 0$.
\end{enumerate}

\section{Visualising Constraints in Parameter Space: Alternative Basis}
\label{app:param_space_alt_basis}

\begin{figure*}
 \centering
    \includegraphics{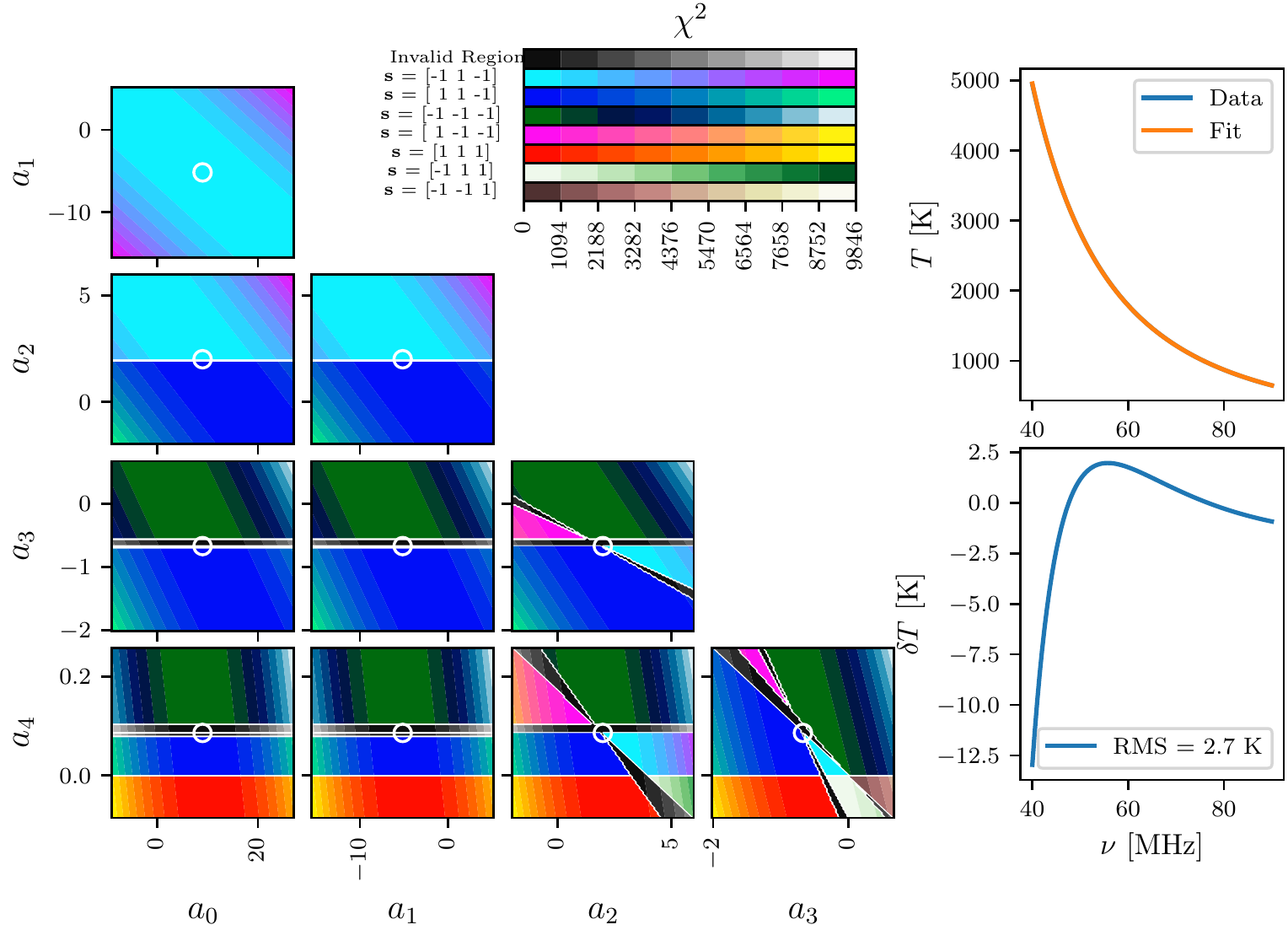}
    \caption{\textbf{Left:} The equivalent of the left panel in \cref{fig:poly_params} using a 5\textsuperscript{th} order MSF of the form given by \cref{eq:loglog_poly} and constrained in $\log_{10}(y) - \log_{10}(x)$ space. As with \cref{fig:poly_params} black regions show regions in which the MSF condition is violated and the coloured regions illustrate sign combinations for which the constraints are upheld. The ranges on the parameters are determined to be $200\%$ on either side of the optimal values from the MSF fit. In each panel two of the parameters are varied while the others are maintained at their optimal values. Here, the regions for which the conditions are violated are narrow and consequently multiple discrete sign spaces are found to produce similar $\chi^2$ values. This strongly suggests that the problem is ill defined and hard to solve using the sign space navigation described in \cref{sec:Eff}. \textbf{Top Right:} The mock 21-cm experiment data and the MSF fit for which the parameter space is analysed. $T$ refers to the averaged sky temperature and $\nu$ to the frequency. \textbf{Bottom Right:} The residuals after subtracting the MSF fit from the data set.}
    \label{fig:loglog_poly_params}
\end{figure*}

\cref{fig:loglog_poly_params} shows the resultant parameter spaces when fitting a 5\textsuperscript{th} order MSF to data of the form $y = x^{-2.5}$ using the logarithmic basis function given by \cref{eq:loglog_poly}. As discussed in \cref{sec:qp} the parameter space presented here is unique to the data set and DCF model used. However, it highlights the importance behind the choice of basis and illustrates the differences in the constraints produced when defining the DCF in a different data spaces.

\section{Standard Derivative Sign Patterns}
\label{app:derivatives}

\begin{figure*}
    \centering
    \includegraphics{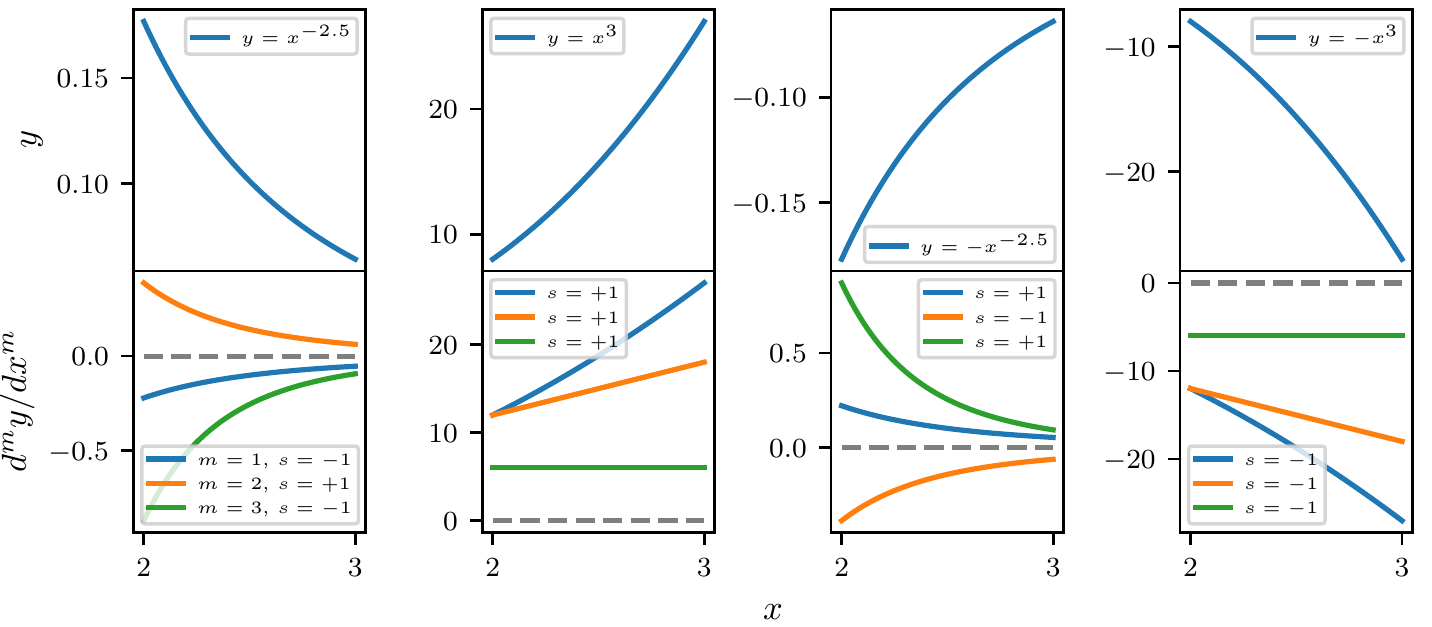}
    \caption{Standard derivative sign patterns associated with four possible standard polynomial data structures. The first row shows example power laws following $y\approx x^{-k}$, $y\approx x^k$, $y\approx -x^{-k}$ and $y\approx -x^k$. The second row shows the derivatives of those power laws up to $m = 3$ and the associated patterns in derivative sign. Note these are not the \maxsmooth~signs and this is discussed in the associated text.}
    \label{fig:standard_der_patterns}
\end{figure*}

In \cref{sec:Eff} we introduce the concept of standard derivative sign patterns for particular polynomial structures. To reiterate and enforce this point \cref{fig:standard_der_patterns} illustrates that the derivatives of a polynomial of the form $y \approx x^k$ are all positive, $y \approx -x^k$ are all negative, $y \approx x^{-k}$ are alternating negative to positive from $m = 1$ and $y \approx -x^{-k}$ are alternating positive to negative from $m = 1$. Since, as discussed in \cref{sec:qp}, \cvxopt~constrains the derivatives, $\mathbf{G}\mathbf{a}$ subject to \cref{eq:cvx_const} we would expect the optimum \maxsmooth~signs for an MSF fit to $y \approx x^k$ to be approximately all negative. Similarly for an MSF fit to a polynomial of the form $y \approx -x^{-k}$ we would expect the optimum signs to be alternating positive to negative for $m \geq 2$.

Note that these standard derivative sign patterns are defined in $y - x$ space. The patterns in $y - z$ space will have similar structures and in logarithmic space they are expected to be different however they will still subscribe to a regular structure.

\bsp	
\label{lastpage}
\end{document}